\documentclass[12pt]{article}
\usepackage{amsmath,amsthm,amsfonts}
\usepackage{fullpage}
\usepackage{graphicx}
\usepackage[dvips]{epsfig}
\usepackage{epsf}
\usepackage{float}
\usepackage{wasysym}

\def\divides{{\  | \ }}
\def\intersect{ \ \cap \ }

\newtheorem{theorem}{Theorem}
\newtheorem{corollary}[theorem]{Corollary}
\newtheorem{proposition}[theorem]{Proposition}
\newtheorem{lemma}[theorem]{Lemma}
\newtheorem{claim}[theorem]{Claim}

\title{Detecting Palindromes, Patterns and Borders in Regular Languages}

\author{Terry Anderson, Narad Rampersad\footnote{Author's current address:
Department of Mathematics and Statistics, University of Winnipeg,
515 Portage Ave., Winnipeg, MB R3B 2E9, Canada.},
Nicolae Santean\footnote{Author's current address:
Department of Computer and Information Sciences,
Indiana University South Bend, 1700 Mishawaka Ave.,
P.O. Box 7111, South Bend, IN 46634, U.S.A.}, and Jeffrey Shallit\\
School of Computer Science\\
University of Waterloo\\
Waterloo, ON  N2L 3G1, Canada\\
{\tt tanderson@uwaterloo.ca} \\
{\tt n.rampersad@uwinnipeg.ca} \\
{\tt nsantean@iusb.edu} \\
{\tt shallit@graceland.uwaterloo.ca}\medskip\\
John Loftus\\
Luzerne County Community College\\
1333 South Prospect Street\\
Nanticoke, PA  18634, U.S.A.\\
{\tt jloftus@luzerne.edu}}

\begin{document}
\date{\today}
\maketitle

\begin{abstract}
Given a language $L$ and a nondeterministic finite automaton $M$,
we consider whether we can determine efficiently (in the size of $M$)
if $M$ accepts at least one word in $L$, or infinitely many words.
Given that $M$ accepts at least one word in $L$,
we consider how long a shortest word can be.
The languages $L$ that we examine include the
palindromes, the non-palindromes, the $k$-powers, the non-$k$-powers,
the powers, the non-powers (also called primitive words), the words matching
a general pattern, the bordered words, and the unbordered words.

\end{abstract}

\section{Introduction}

Let $L \subseteq \Sigma^*$ be a fixed language, and let $M$ be a
deterministic finite automaton (DFA) or nondeterministic finite
automaton (NFA) with input alphabet $\Sigma$.
In this paper we are interested in three questions:

\begin{enumerate}

\item Can we efficiently decide (in terms of the size of $M$)
if $L(M)$ contains at least 
one element of $L$, that is, if $L(M) \intersect L \not= \emptyset$?

\item Can we efficiently decide if $L(M)$ contains infinitely
many elements of $L$, that is, if $L(M) \intersect L$ is infinite?

\item Given that $L(M)$ contains at least one element of $L$, what is
a good upper bound on a shortest element of $L(M) \intersect L$?

\end{enumerate}

We can also ask the same questions about $\overline{L}$, the
complement of $L$.

As an example, consider the case where $\Sigma = \lbrace {\tt a} \rbrace$,
$L$ is the set of primes written in unary, that is,
$\lbrace {\tt a}^i \ : \ i \text{ is prime } \rbrace$, and $M$ is a NFA with
$n$ states.  

To answer questions (1) and (2), we first rewrite $M$ in Chrobak
normal form \cite{Chrobak:1986}.  Chrobak normal form consists of an 
NFA $M'$ with a 
``tail'' of $O(n^2)$ states, followed by a single nondeterministic 
choice to a set of disjoint cycles containing at most $n$ states.  
Computing this normal form can be achieved in $O(n^5)$ steps
by a result of Martinez \cite{Martinez:2002}.  

Now we examine each of the cycles produced by this transformation.
Each cycle accepts a finite union of sets of the form $({\tt a}^t)^*
{\tt a}^c$, where $t$ is the size of the cycle and $c \leq n^2 + n$;
both $t$ and $c$ are given explicitly from $M'$.  Now, by Dirichlet's
theorem on primes in arithmetic progressions, $\gcd(t,c) = 1$ for at
least one pair $(t,c)$ induced by $M'$ if and only if $M$ accepts
infinitely many elements of $L$.  This can be checked in $O(n^2)$
steps, and so we get a solution to question (2) in polynomial time.

Question (1) requires a little more work.  From our answer to question
(2), we may assume that $\gcd(t,c) > 1$ for all pairs $(t,c)$, for
otherwise $M$ accepts infinitely many elements of $L$ and hence at
least one element.  Each element in such a set is of length $kt+c$ for
some $k \geq 0$.   Let $d = \gcd(t,c) \geq 2$.  Then $kt+c = (kt/d +
c/d)d$.  If $k > 1$, this quantity is at least $2d$ and hence
composite.  Thus it suffices to check the primality of $c$ and $t+c$,
both of which are at most $n^2 + 2n$.  We can precompute
the primes $< n^2 + 2n$ in
$O(n^2)$ time using a modification of the sieve of Eratosthenes
\cite{Pritchard:1987}, and check if any of them are accepted.  This
gives a solution to question (1) in polynomial time.

On the other hand, answering question (3) essentially amounts to
estimating the size of the least prime in an arithmetic progression, an
extremely difficult question that is still not fully resolved
\cite{Heath-Brown:1992}, although it is known that there is a
polynomial upper bound.

Even the case where $L$ is regular can be difficult.  Suppose $L$
is represented as the complement of a language accepted by an NFA $M'$ with
$n$ states.  Then if $L(M) =\Sigma^*$, question (1) amounts to asking
if $L(M') \not= \Sigma^*$, which is PSPACE-complete
\cite[Section 10.6]{Aho&Hopcroft&Ullman:1974}.  Question (2) amounts to
asking if $\overline{L(M')}$ is infinite, which is also
PSPACE-complete \cite{Kao&Shallit&Xu:2007}.  
Question (3) amounts to asking for good bounds on the smallest string not
accepted by an NFA.  There is an evident upper bound of $2^n$, and
there are examples known that achieve $2^{cn}$ for some constant
$c > 0$, but more detailed analysis is still lacking
\cite{Ellul&Krawetz&Shallit&Wang:2004}.

Thus we see that asking these questions, even for relatively simple
languages $L$, can quickly take us to
the limits of what is known in formal language theory and number theory.

In this paper we examine questions (1)--(3) in the case where $M$
is an NFA and $L$ is either the set of palindromes, the set of
$k$-powers, the set of powers, the set of words matching a general pattern,
the set of bordered words, or their complements.

   In some of these cases, there is previous work.
For example, Ito et al.\
\cite{Ito&Katsura&Shyr&Yu:1988} studied several
circumstances in which primitive words (non-powers) may appear in
regular languages. As a typical result in
\cite{Ito&Katsura&Shyr&Yu:1988}, we mention:
``A DFA over an alphabet of $2$ or more letters accepts a primitive
word iff it accepts one of length $\leq 3n-3$, where $n$ is the
number of states of the DFA''. 
Horv\'ath, Karhum\"aki and Kleijn \cite{Horvath&Karhumaki&Kleijn:1987} 
addressed the decidability problem of whether a language
accepted by an NFA is palindromic (i.e., every element is a palindrome).
They showed that the
language accepted by an NFA with $n$ states is palindromic
if and only if all its words of length shorter than $3n$
are palindromes. 

     Here is a summary of the rest of the paper.  In section~\ref{nn},
we define the objects of study and our notation.  

In section~\ref{onepal}, we begin our study of palindromes.  We give
efficient algorithms to test if an NFA accepts at least one palindrome,
or infinitely many.  We also show that a shortest palindrome accepted
is of length at most quadratic, and further, that quadratic examples
exist.  In section~\ref{algpal}, we give efficient algorithms to test
if an NFA accepts at least one non-palindrome, or infinitely many.
Further, we give a tight bound on the length of a shortest
non-palindrome accepted.

In section~\ref{pow_test}, we begin our study of patterns.  We show that
it is PSPACE-complete to test if a given NFA accepts a word matching a
given pattern.  As a special case of this problem we consider testing
if an NFA accepts a $k$-power.  We give a 
algorithm to test if a $k$-power is accepted that is polynomial in $k$.
If $k$ is not fixed, the problem is PSPACE-complete.  
We also study the problem of accepting a power of exponent $\geq k$,
and of accepting infinitely many $k$-powers.

In section~\ref{kp},
we give a polynomial-time algorithm to decide if a non-$k$-power
is accepted.  We also give upper and lower bounds
on the length of a shortest $k$-power accepted.  In
section~\ref{powers}, we give an efficient algorithm for
determining if an NFA accepts at least one non-power.
In section~\ref{smallkp}, we bound the length of the smallest power.
Section~\ref{add2pow} gives some additional results on powers.

In section~\ref{bord}, we show how to test if an NFA accepts a bordered
word, or infinitely many,
and show that a shortest bordered word accepted can be of
quadratic length.  In section~\ref{unbord} we give an algorithm
to test if an NFA accepts an unbordered word, or infinitely many,
and we establish a linear upper bound on the length of a shortest
unbordered word.

\section{Notions and notation}\label{nn}

Let $\Sigma$ be an alphabet, i.e., a nonempty, finite set
of symbols (letters). By $\Sigma^*$ we denote the set of
all finite words (strings of symbols) over $\Sigma$, and by
$\epsilon$, the empty word (the word having zero
symbols). The operation of concatenation (juxtaposition) of
two words $u$ and $v$ is denoted by $u\cdot v$, or simply
$uv$.  If $w \in \Sigma^*$ is written in the form $w=xy$ for
some $x,y \in \Sigma^*$, then the word $yx$ is said to be a
{\it conjugate} of $w$.

For $w\in \Sigma^*$, we denote by $w^R$ the word
obtained by reversing the order of symbols in $w$. A {\it
palindrome} is a word $w$ such that $w = w^R$. If $L$ is a
language over $\Sigma$, i.e., $L\subseteq \Sigma^*$, we say
that $L$ is {\it palindromic} if every word $w \in L$ is a
palindrome.

Let $k \geq 2$ be an integer.  A word $y$ is a
\emph{$k$-power} if $y$ can be written as $y = x^k$ for
some non-empty word $x$.  If $y$ cannot be so written for
any $k \geq 2$, then $y$ is \emph{primitive}. A $2$-power
is typically referred to as a \emph{square}, and a
$3$-power as a \emph{cube}.

Patterns are a generalization of powers.  A \emph{pattern}
is a non-empty word $p$ over a \emph{pattern alphabet} $\Delta$.  The
letters of $\Delta$ are called \emph{variables}.  A pattern
$p$ \emph{matches} a word $w \in \Sigma^*$ if there exists a non-erasing
morphism $h : \Delta^* \to \Sigma^*$ such that $h(p) = w$.  Thus,
a word $w$ is a $k$-power if it matches the pattern $a^k$.

      Bordered words are generalizations of powers.  We say a
word $x$ is {\it bordered}
if there exist words $u \in \Sigma^+$, $w \in \Sigma^*$
such that $x = uwu$.  In this case, the word $u$ is said to be a
{\it border} for $x$.  Otherwise, $x$ is {\it unbordered}.

A nondeterministic finite automaton (NFA) over $\Sigma$
is a $5$-tuple $M=(Q, \Sigma, \delta, q_0, F)$ where $Q$
is a finite set of states, $\delta : Q\times \Sigma
\rightarrow 2^{Q}$ is a next-state
function, $q_0$ is an initial state and $F\subseteq Q$ is a
set of final states. We sometimes view $\delta$ as a
transition table, i.e., as a set consisting of tuples $(p,
a, q)$ with $p, q\in Q$ and $a\in\Sigma$.
The machine $M$ is deterministic (DFA) if $\delta$ is a function
mapping $Q\times \Sigma\rightarrow Q$. We consider only {\em
complete} DFAs, that is, those whose transition function
is a total function.  Sometimes we use NFA-$\epsilon$, which are
NFAs that also allow transitions on the empty word.

The size of $M$ is the total number $N$ of its
states and transitions. When we want to emphasize the components of $M$,
we say $M$ has $n$ states and $t$ transitions, and define $N := n+t$.
The language of $M$,
denoted by $L(M)$, belongs to the family of {\em regular
languages} and consists of those words accepted by $M$ in
the usual sense. A {\em successful path}, or {\em
successful computation} of $M$ is any computation starting
in the initial state and ending in a final state. The label
of a computation is the input word that triggered it; thus,
the language of $M$ is the set of labels of all successful
computations of $M$.

A state of $M$ is {\em accessible} if there exists a path in the
associated transition graph, starting from $q_0$ and ending
in that state. By convention, there exists a path from each
state to itself labeled with $\epsilon$. A state $q$ is
{\em coaccessible} if there exists a path from $q$ to some
final state. A state which is both accessible and
coaccessible is called {\em useful}, and if it is not
coaccessible it is called {\em dead}.

We note that if $M$ is an NFA or NFA-$\epsilon$, we can remove all states that
are not useful in linear time (in the number of states and transitions)
using depth-first search.  We observe that $L(M) \not= \emptyset$ if
and only if any states remain after this process, which can be
tested in linear time.  Similarly, if $M$ is a NFA,  then $L(M)$ is infinite
if and only if the corresponding digraph has a directed cycle.  
This can also be tested in linear time.

If $M$ is an NFA-$\epsilon$, then to check if $L(M)$ is infinite
we need to know not only that the corresponding digraph has a cycle, but
that it has a cycle labeled by a non-empty word.  This can also be
checked in linear time as follows.  Let us suppose that all non-useful
states of $M$ have been removed.  We wish to test whether there is
some edge of the digraph of $M$ that is part of some cycle and is not
labeled by the empty word.  We now observe that an edge of a digraph
belongs to a directed cycle if and only if both of its endpoints lie within
the same strongly connected component.  It is well known that the strongly
connected components of a graph can be computed in linear time
(see \cite[Section~22.5]{CLRS01}).  Once the strongly connected components
of the NFA-$\epsilon$ are known, we simply check the edges not
labeled by $\epsilon$ to determine if there is such an edge with both
endpoints in the same strongly connected component.  Thus we can
determine if $L(M)$ is infinite in linear time.

Although the results of this paper are generally stated as applying
to NFA's, by virtue of the preceding algorithm, one sees that the
results apply equally well to NFA-$\epsilon$'s.

We will also need the following well-known results
\cite{Hopcroft&Ullman:1979}:

\begin{theorem}
Let $M$ be an NFA with $n$ states.  Then
\begin{itemize}
\item[(a)]   $L(M) \not= \emptyset$ if and only if $M$ accepts a word
of length $< n$.

\item[(b)] $L(M)$ is infinite if and only if $M$ accepts a word
of length $\ell$, $n \leq \ell < 2n$.
\end{itemize}
\label{hopcroft}
\end{theorem}

If $L \subseteq \Sigma^*$ is a language, the \emph{Myhill--Nerode equivalence
relation} $\equiv_L$ is the equivalence relation defined as
follows:  for $x,y \in \Sigma^*$, $x \equiv_L y$ if for all $z \in \Sigma^*$,
$xz \in L$ if and only if $yz \in L$.  The classical Myhill--Nerode theorem
asserts that if $L$ is regular, the equivalence relation $\equiv_L$ has
only finitely many equivalence classes.

For a background on finite automata and regular languages
we refer the reader to Yu \cite{YU97}.

\section{Testing if an NFA accepts at least one palindrome}
\label{onepal}

     Over a unary alphabet, every string is a palindrome, so problems
(1)--(3) become trivial.  Let us assume, then, that the alphabet $\Sigma$
contains at least two letters.  Although the palindromes over such an
alphabet are not regular, the language
$$ \lbrace x \in \Sigma^* \ : \ x x^R \in  L(M) \text{ or there exists } a \in \Sigma \text{ such that } x a x^R \in L(M) \rbrace$$
is, in fact, regular, as is often shown in a beginning course in formal
languages \cite[p.\ 72, Exercise 3.4 (h)]{Hopcroft&Ullman:1979}.  We
can take advantage of this as follows:

\begin{lemma}
      Let $M$ be an NFA with $n$ states and $t$ transitions.  Then there
exists an NFA-$\epsilon$ $M'$ with $n^2+1$ 
states and $\leq 2t^2$ transitions such that
$$L(M') = \lbrace x \in \Sigma^* \ : \ x x^R \in L(M) \text{ or there
	exists } a \in \Sigma \text{ such that } x a x^R \in L(M) \rbrace.$$
\label{pal-con}
\end{lemma}

\begin{proof}
Let $M=(Q, \Sigma, \delta, q_0, F)$ be an NFA
with $n$ states. We construct an NFA-$\epsilon$
$M' = (Q', \Sigma, \delta', q'_0, F')$ as follows:
We let $Q'=Q\times Q\cup \{q_0'\}$, where $q_0'$ is the new initial state,
and we define 
the set of final states by
$$F' = \lbrace [p, p] \ : \ p\in Q\}\cup \{[p, q] \ : \ \text{ there exists }
a \in \Sigma \text{ such that } q \in \delta(p,a) \rbrace.$$
The transition function $\delta'$ is defined as follows:
$$ \delta'(q'_0, \epsilon) = \lbrace [q_0,q] \ : \ q \in F \rbrace$$
and
$$\delta'([p,q], a) = \lbrace [r,s] \ : \ r \in \delta(p,a) \text{ and }
q \in \delta(s, a) \rbrace.$$

It is clear that $M'$ accepts the desired language and consists of at most
$n^2+1$ states and $2t^2$ transitions.
\end{proof}

\begin{corollary}
    Given an NFA $M$ with $n$ states and $t$ transitions,
we can determine if $M$ accepts a palindrome in $O(n^2 + t^2)$ time.
\end{corollary}

\begin{proof}
      We create $M'$ as in the proof of Lemma~\ref{pal-con},
and remove all states that are not useful, and
their associated transitions.  Now $M$ accepts
at least one palindrome if and only if $L(M')\not=\emptyset$, which can
be tested in time linear in the number of transitions and states of $M'$.
\end{proof}

      From Lemma~\ref{pal-con}, we obtain two other interesting
corollaries.

\begin{corollary}
      Given an NFA $M'$, we can determine if $L(M)$ contains infinitely
many palindromes in quadratic time.
\label{inf-pal}
\end{corollary}

\begin{proof}
       We create $M'$ as in the proof of Lemma~\ref{pal-con}, and remove
all states that are not useful, and their associated transitions.
$M$ accepts infinitely many palindromes if and only if $L(M')$ is infinite,
which can be tested in linear time, as described in Section~\ref{nn}.
\end{proof}

\begin{corollary}
     If an NFA $M$ accepts at least one palindrome, it accepts a
palindrome of length $\leq 2n^2 -1$.
\end{corollary}

\begin{proof}
      Suppose $M$ accepts at least one palindrome.  Then $M'$, as in
Lemma~\ref{pal-con}, accepts a word.  Although $M'$ has $n^2+1$ states,
the only transition from the initial state $q'_0$ is 
an $\epsilon$-transition to one of the other $n^2$ states.  Thus if
$M'$ accepts a word, it must accept a word of length $\leq n^2 - 1$.
Then $M$ accepts
either $w w^R$ or $w a w^R$, and both are palindromes, so $M$
accepts a palindrome of length at most $2(n^2 - 1) + 1 = 2n^2 - 1$.
\end{proof}

     For a different proof of this corollary, see Rosaz \cite{Rosaz:2002}.

      We observe that the quadratic bound is tight, up to 
a multiplicative constant, in the case of alphabets with at
least two letters, and even for DFAs:

\begin{proposition}
For infinitely many $n$ there exists a DFA $M$ with $n$ states
over a $2$-letter alphabet such that
    \begin{itemize}
    \item[(a)] $M$ has $n$ states;
    \item[(b)] The shortest palindrome accepted by $M_n$ is
    of length $\geq n^2/2 - 3n + 5$.
    \end{itemize}
\end{proposition}

\begin{proof}
     For $t \geq 2$,
consider the language $L_t = ({\tt a}^t)^+ {\tt b} ({\tt a}^{t-1})^+$.
This language evidently can be accepted by a DFA with $n = 2t+2$ states.
For a word $w \in L_t$ to be a palindrome, we must have
$w = {\tt a}^{c_1 t} {\tt b} {\tt a}^{c_2 (t-1)}$, for some
integers $c_1, c_2 \geq 1$, with $c_1 t = c_2 (t-1)$.  Since $t$ and
$t-1$ are relatively prime, we must have $t-1 \divides c_1$ and
$t \divides c_2$.  Thus the shortest palindrome in $L_n$ is
${\tt a}^{t(t-1)} {\tt b} {\tt a}^{t(t-1)}$, which is of length
$2t^2 - 2t + 1 = n^2/2 - 3n + 5$.  
\end{proof}

\section{Testing if an NFA accepts at least one non-palindrome}
\label{algpal}

    In this section we consider the problem of deciding if an 
NFA accepts at least one non-palindrome.  Evidently, if an NFA
fails to accept a non-palindrome, it must accept nothing but
palindromes, and so we discuss the opposite decision problem, 

\medskip

\centerline{Given an NFA $M$, is $L(M)$ palindromic?}

\medskip

    Again, the problem is trivial for a unary alphabet, so we 
assume $|\Sigma| \geq 2$.

Horv\'ath, Karhum\"aki, and Kleijn
\cite{Horvath&Karhumaki&Kleijn:1987} proved that the
question is recursively solvable.
In particular, they proved the following theorem:

\begin{theorem}
$L(M)$ is palindromic if and only if $\lbrace x \in L(M) \
: \ |x| < 3n \rbrace$ is palindromic, where $n$ is the
number of states of $M$. \label{hkk}
\end{theorem}

While a naive implementation of Theorem~\ref{hkk} would
take exponential time, in this section we show how to
test palindromicity in polynomial time. We
also show the bound of $3n$ in
Theorem~\ref{hkk} is tight for NFAs, and we improve the bound for
DFAs.

\begin{figure}[H]
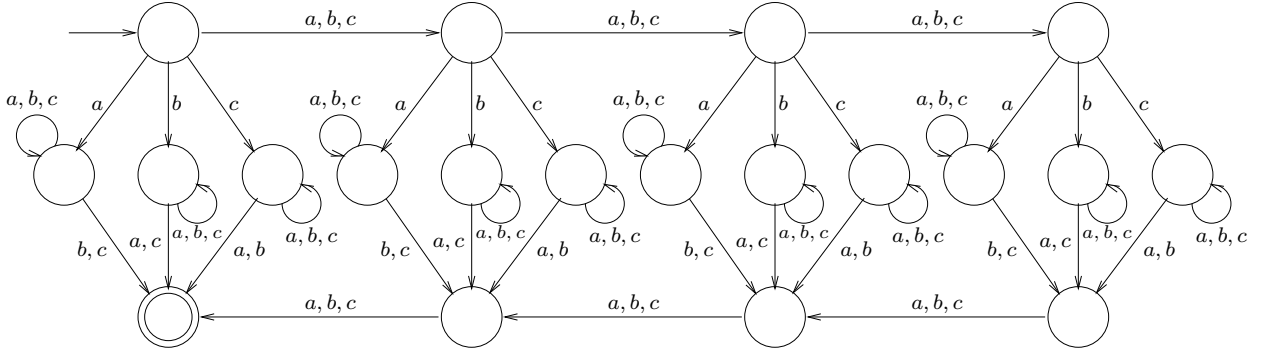

\input nonpal.tex
\caption{Accepting non-palindromes over $\lbrace {\tt
a,b,c} \rbrace$ for $s = 10$.} \label{fig:pred2}
\end{figure}

First, we show how to construct a ``small'' NFA $M'_s$, for
some integer $s >1$, that has the following properties:

\begin{itemize}
\item[(a)] no word in $L(M'_s)$ is a palindrome;

\item[(b)] $M'_s$ accepts all non-palindromes of length $< s$  (in addition to some
other non-palindromes).

\end{itemize}
     The idea in this construction is the following:  on input
$w$ of length $r<s$, we ``guess'' an index $i$, $1 \leq i
\leq r/2$, such that $w[i] \not= w[r+1-i]$.  We then
``verify'' that there is indeed a mismatch $i$ characters
from each end. We can re-use states, as illustrated in
Figure~\ref{fig:pred2} for the case $\Sigma = \lbrace {\tt
a,b,c} \rbrace$ and $s = 10$.

      The resulting NFA $M'_s$ has
$O(|\Sigma| s)$ states
and
$O(|\Sigma|^2 s)$ transitions.  A similar construction appears
in \cite{Shallit&Breitbart:1996}.

Given an NFA $M$ with $n$ states, we now construct the
cross-product with $M'_{3n}$, and obtain an NFA $A$ that
accepts $L(M) \ \cap \ L(M'_{3n})$. We claim that $L(A) =
\emptyset$ if and only if $L(M)$ is palindromic. For if
$L(A) = \emptyset$, then $M$ accepts no non-palindrome of
length $< 3n$, and so by Theorem~\ref{hkk}, $L(M)$ is
palindromic. If $L(A) \not= \emptyset$, then since
$L(M'_{3n})$ contains only non-palindromes, we see that
$L(M)$ is not palindromic.

We can determine if $L(A) = \emptyset$ efficiently by
adding a new final state $q_f$ and
$\epsilon$-transitions from all the final states of $A$
to $q_f$, then performing a depth-first search to detect
whether there are any paths from $q_0$ to $q_f$.  This can
be done in time linear in the number of states and
transitions of $A$.  If $M$ has $n$ states and $t$
transitions, then $A$ has $O(n^2)$ states and
$O(tn)$ transitions.   Hence we have proved the following theorem.

\begin{theorem}
Let $M$ be an NFA with $n$ states and $t$ transitions.
The algorithm sketched above determines whether
$M$ accepts a palindromic language in $O(n^2 + tn)$ time.
\label{thm2}
\end{theorem}

      A different method runs slightly slower, but allows us
to do a little more.    We can mimic the construction for palindromes
in Section~\ref{onepal}, but adapt it for non-palindromes.  Given
an NFA $M$, we construct an NFA-$\epsilon$ $M'$ that accepts the language
\begin{eqnarray*}
\lbrace x \in \Sigma^* &:& \text{there exists } x' \in \Sigma^*,
a \in \Sigma \text{ such that } |x| = |x'|, x \not= {x'}^R,\\
&&\text{ and } x x' \in L(M) \text{ or } x a x' \in L(M) \rbrace.
\end{eqnarray*}
The construction is similar to that in Lemma~\ref{pal-con}.  On input
$x$, we simulate $M$ on $x x'$ and $x a x'$ symbol-by-symbol, moving
forward from the start state and backward from a final state.
We need an additional boolean ``flag'' for each state to record whether or not
we have processed a character in $x'$ that would mismatch the corresponding
character in $x$.   If $M$ has $n$ states and $t$ transitions,
this construction produces an NFA-$\epsilon$ $M'$ with
$\leq 1+2n^2$ states and $O(t^2)$ transitions.  From this we get,
in analogy with Corollary~\ref{inf-pal}, the following proposition.

\begin{proposition}
     Given an NFA $M$ with $n$ states and $t$ transitions, we can determine in 
$O(n^2 + t^2)$ time if $M$ accepts infinitely many non-palindromes.
\end{proposition}

We now turn to the question of the optimality of the $3n$
bound given in Theorem~\ref{hkk}. For an NFA over an
alphabet of at least $2$ symbols, the bound is indeed
optimal, as the following example shows.

\begin{proposition}
Let $\Sigma$ be an alphabet of at least two symbols, containing the
letters $\tt a$ and $\tt b$.
For $n \geq 1$
define $L_n =  ({\tt a}^{n-1} \Sigma)^* {\tt a}^{n-1}$.
Then $L_n$ can be accepted by an NFA with $n$ states and a shortest
non-palindrome in $L_n$ is ${\tt a}^{n-1} {\tt a} {\tt a}^{n-1} {\tt b}
{\tt a}^{n-1}$.
\label{prope}
\end{proposition}

\begin{proof}
The details are straightforward.
\end{proof}

For DFAs, however, the bound of $3n$ can be improved to
$3n-3$. To show this, we first prove the following lemma. A
language $L$ is called {\it slender} if there is a constant
$C$ such that, for all $n \geq 0$, the number of words of
length $n$ in $L$ is less than $C$. The following
characterization of slender regular languages has been
independently rediscovered several times
\cite{Kunze&Shyr&Thierrin:1981,Shallit:1994,Paun&Salomaa:1995}.

\begin{theorem}
\label{slender}
Let $L \subseteq \Sigma^*$ be a regular language.  Then $L$ is slender
if and only if it can be written
as a finite union of languages of the form $u v^* w$, where
$u,v,w \in \Sigma^*$.
\end{theorem}

Next we prove the following useful lemma concerning DFAs accepting
slender languages.

\begin{lemma}
Let $L$ be a slender language accepted by a DFA $M$ with
$n$ states, over an alphabet of two or more symbols.  Then
$M$ must have a dead state. \label{dead-lemma}
\end{lemma}

\begin{proof}
Without loss of generality, assume that every state of $M =
(Q, \Sigma, \delta, q_0, F)$ is reachable from $q_0$, and
that $\Sigma$ contains the symbols $a$ and $b$. We distinguish two
cases:
\begin{enumerate}
\item $M$ accepts a finite language. Consider the states reached
from $q_0$ on $a$, $a^2$, $a^3, \ldots$ Eventually some
state $q$ must be repeated.  This state $q$ must be a dead
state, for if not, $M$ would accept an infinite language.

\item $M$ accepts an infinite language. Then $M$ has at
least one {\em fruitful} cycle, that is, a cycle that
produces infinitely many words in $L(M)$ as labels of
paths starting at $q_0$, entering the cycle, going around
the cycle some number of times,  then exiting and
eventually reaching a final state. Let $C_1$ be one
fruitful cycle, and consider the following successful path
involving $C_1$: $q_0 {\buildrel\alpha\over\longrightarrow}
q {\buildrel u \over\longrightarrow} q {\buildrel \beta
\over\longrightarrow} f$, where $f\in F$ and the repetition
of $q$ denotes the cycle $C_1$, labeled with $u$. Without
loss of generality assume the first letter of $u$ is $a$.
Since $M$ is complete, denote $p=\delta(q, b)$.

We claim that from $p$ one cannot reach a fruitful cycle $C_2$.
Indeed, let's assume the contrary; this means that there exists
a successful path $q_0 {\buildrel\alpha\over\longrightarrow}
q {\buildrel u \over\longrightarrow} q {\buildrel \gamma
\over\longrightarrow} r {\buildrel v \over\longrightarrow}
r {\buildrel \mu \over\longrightarrow f'}$, with $f'\in F$
and the repetition of $r$ denotes the cycle $C_2$ labeled
with $v$. Let $n$ be an arbitrary integer, and $0 \leq i \leq n$.
There exist two integers $k, l$ such that
$k|u|=l|v|=m$. With this notation, observe that the
words $\alpha u^{k(n-i)}\gamma v^{l(n+i)}\mu$ are all
accepted by $M$ and have the same length $2mn + |\alpha\gamma\mu|$.
Since there are $n+1$
such words, this proves that $L(M)$ has $\Omega(n)$ words of length $n$
for large $n$---a contradiction.

Thus, there exist a finite number of successful paths
starting from $p$. However, considering the states reached
from $p$ by the words $a$, $a^2$, $a^3, \ldots$, one such
state must repeat. This state is dead, for the alternative would
contradict the finiteness of successful paths from $p$.
\end{enumerate}
\end{proof}

\begin{corollary}
    If $M$ is a DFA over an alphabet of at least two letters
and $L(M)$ is palindromic, then $M$ has a dead state.
\label{dead-cor}
\end{corollary}

\begin{proof}
    If $L(M)$ is palindromic, then by
\cite[Theorem 8]{Horvath&Karhumaki&Kleijn:1987}
it can be written as a finite union of languages of the form
$u v (tv)^* u^R$, where $u, v, t \in \Sigma^*$ and $v, t$ are
palindromes.  By Theorem~\ref{slender}, this means
$L(M)$ is slender.  By Lemma~\ref{dead-lemma}, $M$ has a dead state.
\end{proof}

   We are now ready to prove the improved bound of $3n-3$ for DFAs.

\begin{theorem}
Let $M$ be a DFA with $n$ states.  Then $L(M)$ is palindromic if and
only if $\lbrace x \in L(M) \ : \ |x| < 3n-3 \rbrace$ is palindromic.
\end{theorem}

\begin{proof}
      One direction is clear.

     If $M = (Q, \Sigma, \delta, q_0, F)$ is over a unary alphabet,
then $L(M)$ is always palindromic, so the criterion is trivially true.

    Otherwise $M$ is over an alphabet of at least two letters.
Assume  $\lbrace x \in L(M) \ : \ |x| < 3n-3 \rbrace$ is palindromic.  From
Corollary~\ref{dead-cor}, we see that $M$ must have a dead state.
But then we can delete such a dead state and all associated transitions,
and all states reachable from the deleted dead state, to get a new NFA $M'$
with at most $n-1$ states that accepts the same language.
We know from Theorem~\ref{hkk} that the palindromicity of 
$\lbrace x \in L(M') \ : \ |x| < 3n-3 \rbrace$ implies that
$M'$ is palindromic.
\end{proof}

   Finally, we observe that $3n-3$ is the best possible bound
in the case of DFAs.  To do so, we simply use the language $L_n$
from Proposition~\ref{prope} and observe it can be accepted by
a DFA with $n+1$ states; yet the shortest non-palindrome is of
size $3n-1$.

We end this section by noting that the related, but fundamentally
different, problem of testing if $L = L^R$ was shown by Hunt
\cite{Hunt:1973} to be PSPACE-complete.

\section{Testing if an NFA accepts a word matching a pattern}
\label{pow_test}

In this section we consider the computational complexity of testing
if an NFA accepts a word matching a given pattern.
Specifically, we consider the following decision problem.

\begin{quotation}
\noindent{\bf NFA PATTERN ACCEPTANCE}

\noindent INSTANCE: An NFA $M$ over the alphabet $\Sigma$ and a
pattern $p$ over some alphabet $\Delta$.

\noindent QUESTION: Does there exist $x \in \Sigma^+$ such that
$x \in L(M)$ and $x$ matches $p$?
\end{quotation}

Since the pattern $p$ is given as part of the input, this problem
is actually somewhat more general than the sort of problem
formulated as Question~1 of the introduction, where the language
$L$ was fixed.

We first consider the following result of Restivo and Salemi
\cite{Restivo&Salemi:2001} (a more detailed proof appears in
\cite{Castiglione&Restivo&Salemi:2004}).  We give here a boolean matrix
based proof (see Zhang \cite{Zhang:1999} for a study of this boolean matrix
approach to automata theory) that illustrates our general approach to
the other problems treated in this section.

\begin{theorem}[Restivo and Salemi]
\label{res_sal}
Let $L$ be a regular language and let $\Delta$ be an alphabet.
The set $P_\Delta$ of all non-empty patterns $p \in \Delta^*$
such that $p$ matches a word in $L$ is effectively regular.
\end{theorem}

\begin{proof}
Let $M = (Q,\Sigma,\delta,q_0,F)$ be an NFA such that $L(M) = L$.
Suppose that $Q = \{0,1,\ldots,n-1\}$.
For $a \in \Sigma$, let $B_a$ be the $n \times n$ boolean matrix whose
$(i,j)$ entry is $1$ if $j \in \delta(i,a)$ and $0$ otherwise.
Let $\mathcal{B}$ denote the semigroup generated by the $B_a$'s
along with the identity matrix.
For $w = w_0 w_1 \cdots w_s$, where $w_i \in \Sigma$ for $i = 0,\ldots,s$,
we write $B_w$ to denote the matrix product $B_{w_0} B_{w_1} \cdots B_{w_s}$.

Without loss of generality, let $\Delta = \{1,2,\ldots,k\}$.
Observe that there exists a non-empty pattern
$p = p_0 p_1 \cdots p_r$, where $p_i \in \Delta$ for $i = 0,\ldots,r$,
and a non-erasing morphism $h : \Delta^* \to \Sigma^*$ such that $h(p) \in L$
if and only if there exist $k$ boolean matrices
$B_1,\ldots,B_k \in \mathcal{B}$ such that $B_i = B_{h(i)}$ for
$i \in \Delta$ and $B = B_{p_0} B_{p_1} \cdots B_{p_r}$ describes
an accepting computation of $M$.

We construct an NFA $M' = (Q',\Delta,\delta',P,F')$ for $P_\Delta$
as follows.  For simplicity, we permit $M'$ to have multiple initial states,
as specified by the set $P$.  We define $Q' = \mathcal{B}^{k+1}$.
The set $P$ of initial states is given by $P = \mathcal{B}^k \times I$,
where $I$ denotes the identity matrix.  In other words, the NFA $M'$ uses the
first $k$ components of its state to record an initial guess of $k$ boolean
matrices $B_1,\ldots,B_k \in \mathcal{B}$.  Let $[B_1,\ldots,B_k,A]$
denote some arbitrary state of $M'$.  For $i=1,\ldots,k$, the
transition function $\delta'$ maps $[B_1,\ldots,B_k,A]$ to
$[B_1,\ldots,B_k,AB_i]$.  In other words, on input
$p = p_0 p_1 \cdots p_r\in \Delta^*$, $M'$ uses the last component of
its state to compute the product $B = B_{p_0} B_{p_1} \cdots B_{p_r}$.
The set $F'$ of final states of $M'$ consists of all states of the form
$[B_1,\ldots,B_k,B]$, where the matrix $B$ contains a $1$ in some entry
$(0,j)$, where $j \in F$.  In other words, $M'$ accepts if and only if
$B$ describes an accepting computation of $M$.
\end{proof}

By consider unary patterns of the form $a^k$, we obtain the following
corollary of Theorem~\ref{res_sal}.

\begin{corollary}
Let $L \subseteq \Sigma^*$ be a regular language.  The set of exponents $k$
such that $L$ contains a $k$-power is the union of a finite set with a finite
union of arithmetic progressions.  Further, this set of exponents is
effectively computable.
\end{corollary}

Observe that Theorem~\ref{res_sal} implies the decidability
of the {\bf NFA PATTERN ACCEPTANCE} problem.  We prove the following
stronger result.

\begin{theorem}
\label{pattern}
The {\bf NFA PATTERN ACCEPTANCE} problem is PSPACE-complete.
\end{theorem}

\begin{proof}
We first show that the problem is in PSPACE.  By Savitch's theorem
\cite{Savitch:1970} it suffices to give an NPSPACE algorithm.
Let $M = (Q,\Sigma,\delta,q_0,F)$, where $Q = \{0,1,\ldots,n-1\}$.
For $a \in \Sigma$, let $B_a$ be the $n \times n$ boolean matrix whose
$(i,j)$ entry is $1$ if $j \in \delta(i,a)$ and $0$ otherwise.
Let $\mathcal{B}$ denote the semigroup generated by the $B_a$'s along with
the identity matrix.  For $w = w_0 w_1 \cdots w_s \in \Sigma^*$, we write
$B_w$ to denote the matrix product $B_{w_0} B_{w_1} \cdots B_{w_s}$.

Let $\Delta$ be the set of letters occuring in $p$.  We may suppose that
$\Delta = \{1,2,\ldots,k\}$.  First, we non-deterministically guess
$k$ boolean matrices $B_1, \ldots, B_k$.  Next, for each $i$, we
verify that $B_i$ is in the semigroup $\mathcal{B}$ by
non-deterministically guessing a word $w = w_0 w_1 \cdots w_s$
such that $B_i = B_w$.  Since there are at most
$2^{n^2}$ possible $n \times n$ boolean matrices, we may assume that
$s \leq 2^{n^2}$.  We thus guess $w$ symbol-by-symbol and compute a
sequence of matrices
\[
B_{w_1}, B_{w_1w_2}, \ldots, B_{w_1w_2 \cdots w_s},
\]
reusing space after perfoming each matrix multiplication.
We maintain an $O(n^2)$ bit counter to keep track of the
length $s$ of our guessed word $w$.  If $s$ exceeds $2^{n^2}$, we reject
on this branch of the non-deterministic computation.

Finally, if $p = p_0 p_1 \cdots p_r$, we compute the matrix product
$B = B_{p_0} B_{p_1} \cdots B_{p_r}$ and accept if and only if $B$
describes an accepting computation of $M$.

To show hardness we reduce from the following PSPACE-complete problem 
\cite[Problem AL6]{Garey&Johnson:1979}.

\begin{quotation}
\noindent{\bf DFA INTERSECTION}

\noindent INSTANCE: An integer $k \geq1$ and $k$ DFAs
$A_1,A_2,\ldots,A_k$, each over the alphabet $\Sigma$.

\noindent QUESTION: Does there exist $x \in \Sigma^*$ such that $x$
is accepted by each $A_i$, $1 \leq i \leq k$?\qed
\end{quotation}

Let $\#$ be a symbol not in $\Sigma$.  We construct, in linear time, a
DFA $M$ to accept the language
$L(A_1) \,\#\, L(A_2) \,\#\, \cdots L(A_k) \,\#$.
Any word in $L(M)$ matching the pattern $a^k$ is of the form $(x\#)^k$.
It follows that $M$ accepts a word matching $a^k$ if and only if there
exists $x$ such that $x \in L(A_i)$ for $1 \leq i \leq k$.
This completes the reduction.
\end{proof}

We may define various variations or special cases of the
{\bf NFA PATTERN ACCEPTANCE} problem, such as:
{\bf NFA ACCEPTS A $k$-POWER},
{\bf NFA ACCEPTS A $\geq k$-POWER},
{\bf NFA ACCEPTS INFINITELY MANY $k$-POWERS},
{\bf NFA ACCEPTS INFINITELY MANY $\geq k$-POWERS}, etc.
We define and consider the computational complexity of these variations
below.

\begin{quotation}
\noindent{\bf NFA ACCEPTS A $k$-POWER}.

\noindent INSTANCE: An NFA $M$ over the alphabet $\Sigma$ and an
integer $k \geq 2$.

\noindent QUESTION: Does there exist $x \in \Sigma^+$ such that
$M$ accepts $x^k$?
\end{quotation}

\begin{quotation}
\noindent{\bf NFA ACCEPTS A $\geq k$-POWER}.

\noindent INSTANCE: An NFA $M$ over the alphabet $\Sigma$.

\noindent QUESTION: Does there exist $x \in \Sigma^+$ and an integer
$\ell \geq k$ such that $M$ accepts $x^\ell$?
\end{quotation}

    The {\bf NFA ACCEPTS A $\geq k$-POWER} problem is actually an infinite
family of problems, each indexed by an integer $k \geq 2$.
If $k$ is fixed, the {\bf NFA ACCEPTS A $k$-POWER} problem can
be solved in polynomial time, as we now demonstrate.

\begin{proposition}\label{fixed-k}
Let $M$ be an NFA with $n$ states and $t$ transitions, and set
$N = n+t$, the size of $M$.
For any fixed integer $k \geq 2$, there is an algorithm running in
$O(n^{2k-1} t^k) = O(N^{2k-1})$ time
to determine if $M$ accepts a $k$-power.
\end{proposition}

\begin{proof}
For a language $L \subseteq \Sigma^*$, we define
$$L^{1/k} = \{ x \in \Sigma^* : x^k \in L \}.$$

Let $M = (Q, \Sigma, \delta, q_0, F)$ be an NFA with $n$ states.  
We will construct an NFA-$\epsilon$ $M'$ such that $L(M') = L(M)^{1/k}$.
To determine whether or not $M$ accepts a $k$-power, 
it suffices to check whether or not $M'$ accepts a non-empty word.

The idea behind the construction of $M'$ is as follows.
On input $x$, $M'$ first guesses $k-1$ states
$g_1, g_2, \ldots, g_{k-1} \in Q$ and then checks that
\begin{itemize}
\item
  $g_1 \in \delta(q_0,x)$,
\item
  $g_{i+1} \in \delta(g_i,x)$ for $i = 1,2,\ldots,k-2$, and
\item
  $\delta(g_{k-1},x) \cap F \neq \emptyset$.
\end{itemize}
It is clear that such states $g_1, g_2, \ldots, g_{k-1}$ exist
if and only if $x^k \in L(M)$.

Formally, the construction of $M'$ is as follows.
We define the NFA $M' = (Q', \Sigma, \delta', q_0', F')$ such that:

\begin{itemize}
\item $Q' = \{ q_0' \} \cup Q^{2k-1}$.  That is, except for $q_0'$,
each state of $M'$ is a $(2k-1)$-tuple of the form
$[g_1, g_2, \ldots, g_{k-1}, p_0, p_1, \ldots, p_{k-1}]$.
The state $g_i$ represents the $i$-th state guessed from $M$.
The NFA $M'$ will simulate in parallel the computations of $M$ on input
$x$ starting from states $q_0, g_1, g_2, \ldots, g_{k-1}$ respectively.
The state $p_0$ represents the current state of the simulation beginning
from state $q_0$, and the states $p_1, p_2, \ldots, p_{k-1}$ represent the
current states of the simulations beginning from states
$g_1, g_2, \ldots, g_{k-1}$, respectively.

\item $q_0'$ is an additional state not in $Q^{2k-1}$.
This state will have outgoing $\epsilon$-transitions for each
different combination of guesses $g_i$.
The transition function on the start state is defined as
$$\delta'( q_0', \epsilon) = \{ [g_1, g_2, \ldots, g_{k-1},
q_0, g_1, g_2, \ldots, g_{k-1}] : 
\forall i \in \{1,2,\ldots,k-1\}, g_i \in Q \}.$$

\item We define the transition function $\delta'$ on all other states
as:
\begin{eqnarray*}
\lefteqn{\delta'([g_1, g_2, \ldots, g_{k-1}, p_0, p_1, \ldots, p_{k-1}], a)=}\\
& & \{[g_1, g_2, \ldots, g_{k-1}, p_0', p_1', \ldots, p_{k-1}'] :
\forall i \in \{0,1,\ldots,k-1\}, p_i' \in \delta (p_i, a)\}
\end{eqnarray*}
for all $a \in \Sigma$.

\item $F' = \{ [g_1, g_2, \ldots, g_{k-1}, g_1, g_2, \ldots, g_{k-1}, t]: t \in F \}$.
That is, we reach a state in $F'$ on input $x$ exactly when the guessed
states $g_i$ verify the conditions described above.
\end{itemize}

It should be clear from the construction that $M'$ accepts $L(M)^{1/k}$.
The number of states in $M'$ is $n^{2k-1} + 1$, as, except for $q_0'$,
each state is a $(2k-1)$-tuple in which each coordinate can take on
$|Q|=n$ possible values.  For each state there are at most $t^k$ distinct
transitions.  Testing whether or not $L(M')$ accepts a non-empty
word can be done in linear time (since the only $\epsilon$-transitions are
transitions outgoing from $q_0'$), so the running time of our algorithm is
$O(n^{2k-1} t^k)$.
\end{proof}

    As before, we can use the same automaton to test if infinitely many
$k$-powers are accepted.

\begin{corollary}
     We can decide if an NFA $M$ with $n$ states and $t$ transitions
accepts infinitely many $k$-powers in $O(n^{2k-1} t^k)$ time.
\end{corollary}

If $k$ is not fixed, we have the following result, which is an immediate
consequence of Theorem~\ref{pattern} if $k$ is given in unary.  However,
the problem remains in PSPACE even if $k$ is given in binary, as we now
demonstrate.
     
\begin{theorem}
\label{kpow_alg}
The problem {\bf NFA ACCEPTS A $k$-POWER} is PSPACE-complete.
\end{theorem}

\begin{proof}
We first show that the problem is in PSPACE.  By Savitch's theorem
\cite{Savitch:1970} it suffices to give an NPSPACE algorithm.
Let $M = (Q,\Sigma,\delta,q_0,F)$, where $Q = \{0,1,\ldots,n-1\}$.
For $a \in \Sigma$, let $B_a$ be the $n \times n$ boolean matrix whose
$(i,j)$ entry is $1$ if $j \in \delta(i,a)$ and $0$ otherwise.
Let $\mathcal{B}$ denote the semigroup generated by the $B_a$'s.

We non-deterministically guess a boolean matrix $B$ and
verify that $B \in \mathcal{B}$ (i.e., $B = B_x$ for some $x \in \Sigma^*$),
as illustrated in the proof of Theorem~\ref{pattern}.
Finally, we compute $B_x^k$ efficiently by repeated squaring
and verify that $B_x^k$ contains a $1$ in  position $(q_0,f)$ for some
$f \in F$.

The proof for PSPACE-hardness is precisely that given in the proof
of Theorem~\ref{pattern}.
\end{proof}

\begin{theorem}
\label{pow_alg}
For each integer $k \geq 2$, the problem {\bf NFA ACCEPTS A $\geq k$-POWER}
is PSPACE-complete.
\end{theorem}

\begin{proof}
To show that the problem is in PSPACE, we use the same algorithm
as in the proof of Theorem~\ref{kpow_alg}, with the following modification.
In order to verify that $M$ accepts an $\ell$-power for some $\ell \geq k$,
we first observe that by the same argument as in the proof of
Proposition~\ref{exponent_bd} below, if $M$ accepts such an $\ell$-power,
then $M$ accepts an $\ell$-power for $k \leq \ell < k+n$.
Thus, after non-deterministically computing $B_x$, we must compute
$B_x^\ell$ for all $k \leq \ell < k+n$, and verify that at least
one $B_x^\ell$ contains a $1$ in position $(q_0,f)$ for some $f \in F$.

To show PSPACE-hardness, we again reduce from the {\bf DFA INTERSECTION}
problem.  Suppose that we are given $r$ DFAs $A_1,A_2,\ldots,A_r$ and we wish
to determine if the $A_i$'s accept a common word $x$.  We may suppose
that $r \geq k$, since for any fixed $k$ such a restriction does not affect
the PSPACE-completeness of the {\bf DFA INTERSECTION} problem.
Let $j$ be the smallest non-negative integer such that $r+j$ is prime.
By Bertrand's Postulate \cite[Theorem~418]{Hardy&Wright:1979},
we may take $j \leq r$.  We now construct, in linear time,
a DFA $M$ to accept the language
$L(A_1) \,\#\, L(A_2) \,\#\, \cdots L(A_r) \,\# (\Sigma^* \,\#)^j$.
The DFA $M$ accepts a $\geq k$-power if and only if it accepts an
$(r+j)$-power.  Moreover, $M$ accepts an $(r+j)$-power if and only if
there exists $x$ such that $x \in L(A_i)$ for $1 \leq i \leq r$.
This completes the reduction.
\end{proof}

In a similar fashion, we now show that the following decision problems
are PSPACE-complete:

\begin{quotation}
\noindent{\bf NFA ACCEPTS INFINITELY MANY $k$-POWERS}.

\noindent INSTANCE: An NFA $M$ over the alphabet $\Sigma$ and an
integer $k \geq 2$.

\noindent QUESTION: Does $M$ accept $x^k$ for infinitely many words $x$?
\end{quotation}

\begin{quotation}
\noindent{\bf NFA ACCEPTS INFINITELY MANY $\geq k$-POWERS}.

\noindent INSTANCE: An NFA $M$ over the alphabet $\Sigma$.

\noindent QUESTION: Are there infinitely many pairs $(x,i)$ such that
$i \geq k$ and $M$ accepts $x^i$?
\end{quotation}

      Again, the {\bf NFA ACCEPTS INFINITELY MANY $\geq k$-POWERS}
problem is actually an infinite family of problems, each indexed by
an integer $k \geq 2$.
We will prove that these decision problems are PSPACE-complete
by reducing from the following problem.

\begin{quotation}
\noindent{\bf INFINITE CARDINALITY DFA INTERSECTION}.

\noindent INSTANCE: An integer $k \geq 1$ and  $k$ DFAs
$A_1,A_2,\ldots,A_k$, each over the alphabet $\Sigma$.

\noindent QUESTION: Do there exist infinitely many
$x \in \Sigma^*$ such that $x$
is accepted by each $A_i$, $1 \leq i \leq k$?
\end{quotation}

\begin{lemma}
     The decision problem {\bf INFINITE CARDINALITY DFA INTERSECTION}
is PSPACE-complete.
\end{lemma}

\begin{proof}
     First, let's see that the problem is in PSPACE.  If the largest
DFA has $n$ states, then there is a DFA with at most $n^k$ states
that accepts $\bigcap_{1 \leq i \leq k}  L(A_i)$.  Now from
Theorem~\ref{hopcroft} (b), we know that there exist infinitely many
$x$ accepted by each $A_i$ if and only if there is a word $x$
length $\ell$, $n^k \leq \ell < 2n^k$, accepted by all the $A_i$.
We can simply guess the symbols of $x$, ensuring with a counter that
$n^k \leq |x| < 2n^k$, and checking by simulation that $x$ is accepted
by all the $A_i$.  The counter uses at most $k \log n + \log 2$ bits,
which is polynomial in the size of the input.  This shows the problem
is in nondeterministic polynomial space, and hence, by Savitch's theorem
\cite{Savitch:1970}, in PSPACE.

     Now, to see that {\bf INFINITE CARDINALITY DFA INTERSECTION}
is PSPACE-hard, we reduce from {\bf DFA INTERSECTION}.  For each
DFA $A_i = (Q_i, \Sigma, \delta_i, q_{0,i}, F_i)$,
we modify it to $B_i$ as follows:  we add a new initial
state $q'_{0,i}$, and add the same transitions from it as from $q_{0,i}$.
We then change all final states to non-final, and we make $q'_{0,i}$
final.  
We add a transition from all states that were
previously final on a new letter $\cent$ (the same letter is used for
each $A_i$), and a transition from all other states on $\cent$ to a new
dead state $d$.  Finally, we add transitions on all letters from $d$ 
to itself.   We claim $B_i$ is a DFA and
$L(B_i) = (L(A_i)\cent)^*$.   Furthermore,
$\bigcap_{1 \leq i \leq k}  L(A_i) \not= \emptyset$ if and only if
$\bigcap_{1 \leq i \leq k} L(B_i)$ is infinite.

Suppose $\bigcap_{1 \leq i \leq k}  L(A_i) \not= \emptyset$.  Then
there exists $x$ accepted by each of the $A_i$.  Then $(x\cent)^*$ 
is accepted by each of the $B_i$, so 
$\bigcap_{1 \leq i \leq k}  L(B_i) $ is infinite.

Now suppose $\bigcap_{1 \leq i \leq k}  L(B_i) $ is infinite.  Choose
any nonempty 
$x \in \bigcap_{1 \leq i \leq k}  L(B_i)  = \bigcap_{1 \leq i \leq k}
(L(A_i)\cent)^*$.  Thus $x$ must be of the form $y_1 \cent y_2 \cent
\cdots y_j \cent$
for some $j \geq 1$, where each $y_i$ is accepted by all the $A_i$.  
Hence, in particular, $y_1$ is accepted by all the $A_i$, and so
$\bigcap_{1 \leq i \leq k}  L(A_i) \not= \emptyset$.
\end{proof}

We are now ready to prove

\begin{theorem}  
The decision problem {\bf NFA ACCEPTS INFINITELY MANY $k$-POWERS} is
PSPACE-complete.
\end{theorem}

\begin{proof}
    First, let's see that the problem is in PSPACE.  We claim that
an NFA $M$ with $n$ states accepts infinitely many $k$-powers if and only
if it accepts a $k$-power $x^k$ with $2^{n^2} \leq |x| < 2^{n^2+1}$.

One direction is clear.  For the other direction,
we use boolean matrices, as in the proof of Theorem~\ref{kpow_alg}.  
We can construct a DFA $M' = (Q', \Sigma, \delta', q'_0, F')$
of $2^{n^2}$ states that accepts $L^{1/k} =
\lbrace x \in \Sigma^* \ : \ x^k \in L(M) \rbrace$, as follows:  the states are
$n \times n$ boolean matrices.  The initial state $q'_0$ is the
identity matrix. If $B_a$ is the boolean matrix with
a $1$ in entry $(i,j)$ if $j \in \delta(q_i,a)$ and $0$ otherwise,
then $\delta'(B, a) = B B_a$.  The set of final states is
$F' = \lbrace B \ : \ \text{the $(0,j)$ entry of $B^k$ is $1$ for some }
q_j \in F \rbrace.$  

The idea of this construction is that  if $x = a_1 a_2 \cdots a_i$, then
$\delta(q'_0, x) = B_{a_1} \cdots B_{a_i}$.    Now we use
Theorem~\ref{hopcroft} (b) to conclude that $M'$ accepts infinitely
many words if and only if it accepts a word $x$ with
$2^{n^2} \leq |x| < 2^{n^2+1}$.  But $L(M') = L(M)^{1/k}$.

Thus, to check if $M$ accepts infinitely many $k$-powers, we simply
guess the symbols of $x$, stopping when $2^{n^2} \leq |x| < 2^{n^2+1}$,
and verify that $M$ accepts $x^k$.  We can do this by accumulating
$B_{a_1} \cdots B_{a_k}$ and raising the result to the $k$-th power, as before.
We need $n^2+1$ bits to keep track of the counter, so the result is in
NPSPACE, and hence in PSPACE.

Now we argue that {\bf NFA ACCEPTS INFINITELY MANY $k$-POWERS} is
PSPACE-hard.  To do so, we reduce from 
{\bf INFINITE CARDINALITY DFA INTERSECTION}.  Given DFAs
$A_1, A_2, \ldots, A_k$, we can easily construct a DFA $A$ to accept
$L(A_1) \# \cdots L(A_k) \#$.  Clearly $A$ accepts infinitely many
$k$-powers if and only if $\bigcap_{1 \leq i \leq k} L(A_i)$ is infinite.
\end{proof}

\begin{theorem}  
For each integer $k \geq 2$, the problem
{\bf NFA ACCEPTS INFINITELY MANY $\geq k$-POWERS} is PSPACE-complete.
\end{theorem}

\begin{proof}
Left to the reader.
\end{proof}

\section{Testing if an NFA accepts a non-$k$-power}
\label{kp}

In the previous section we showed that it is computationally hard
to test if an NFA accepts a $k$-power (when $k$ is not fixed).
In this section we show how to test if an NFA accepts a
non-$k$-power.  Again, we find it more congenial to discuss the
opposite problem, which is whether an NFA accepts nothing but
$k$-powers.

First, we need several classical
results from the theory of combinatorics on words.  
The following theorem is due to Lyndon and Sch\"utzenberger
\cite{Lyndon&Schutzenberger:1962}.

\begin{theorem}
\label{ls_eqn}
If $x$, $y$, and $z$ are words satisfying an equation $x^i y^j = z^k$,
where $i,j,k \geq 2$, then they are all powers of a common word.
\end{theorem}

The next result is also due to Lyndon and Sch\"utzenberger.

\begin{theorem}
\label{lyn_schu}
Let $u$ and $v$ be non-empty words.  If $uv = vu$, then there exists
a word $x$ and integers $i,j \geq 1$, such that $u = x^i$ and $v = x^j$.
In other words, $u$ and $v$ are powers of a common word.
\end{theorem}

The following result can be derived from Theorem~\ref{lyn_schu}.

\begin{corollary}
\label{ls_cor}
Let $u$ and $v$ be non-empty words.  If $u^r = v^s$ for some $r,s \geq 1$,
then $u$ and $v$ are powers of a common word.
\end{corollary} 

Ito, Katsura, Shyr, and Yu \cite{Ito&Katsura&Shyr&Yu:1988}
gave a proof of the next proposition.

\begin{proposition}
\label{shyr}
Let $u$ and $v$ be non-empty words.  If $u$ and $v$ are not powers of
a common word, then for any integers $r,s \geq 1$, $r \neq s$,
at least one of $u^rv$ or $u^sv$ is primitive.
\end{proposition}

The next result is due to Shyr and Yu \cite{Shyr&Yu:1994}.

\begin{theorem}
\label{p+q+}
Let $p$ and $q$ be primitive words, $p \neq q$.  The set $p^+q^+$
contains at most one non-primitive word.
\end{theorem}

Next we prove the following analogue
of Theorem~\ref{hkk}, from which we will derive an
efficient algorithm for testing if a finite automaton
accepts only $k$-powers.

\begin{theorem}
\label{k-pow}
Let $L$ be accepted by an $n$-state NFA $M$ and let $k \geq 2$ be an integer.
\begin{enumerate}
\item Every word in $L$ is a $k$-power if and only if every word in the set
$\lbrace x \in L : |x| \leq 3n \rbrace$ is a $k$-power.
\item All but finitely many words in $L$ are $k$-powers if and only if
every word in the set $\lbrace x \in L : n \leq |x| \leq 3n \rbrace$
is a $k$-power.
\end{enumerate}
Further, if $M$ is a DFA over an alphabet of size $\geq 2$, then the bound $3n$
may be replaced by $3n-3$.
\end{theorem}

Ito, Katsura, Shyr, and Yu \cite{Ito&Katsura&Shyr&Yu:1988}
proved a similar result for primitive words: namely, that
if $L$ is accepted by an $n$-state DFA over an alphabet of
two or more letters and contains a
primitive word, then it contains a primitive word of length
$\leq 3n-3$. In other words, every word in $L$ is a power if and only if
every word in the set $\lbrace x \in L : |x| \leq 3n-3 \rbrace$ is a power.
However, this result does not imply
Theorem~\ref{k-pow}, as one can easily construct a regular
language $L$ where every word in $L$ that is not a
$k$-power is nevertheless non-primitive:  for example, $L = \lbrace a^{k+1}
\rbrace$.  

We shall use the next result to characterize those regular languages
consisting only of $k$-powers.

\begin{proposition}
\label{prim} Let $u$, $v$, and $w$ be words, $v \neq
\epsilon$, $uw \neq \epsilon$, and let $f,g \geq 1$ be integers, $f \neq g$.
If $uv^fw$ and $uv^gw$ are non-primitive, then $uv^nw$ is
non-primitive for all integers $n \geq 1$. Further, if
$uvw$ and $uv^2w$ are $k$-powers for some integer $k \geq
2$, then $v$ and $uv^nw$ are $k$-powers for all integers $n
\geq 1$.
\end{proposition}

\begin{proof}
Suppose $uv^fw$ and $uv^gw$ are non-primitive.  Then $v^fwu$ and
$v^gwu$ are non-primitive. Let $x$ and $y$ be the primitive roots of $v$ and
$wu$, respectively, so that $v = x^i$ and $wu = y^j$ for some integers
$i,j \geq 1$.  If $x \neq y$, then by Proposition~\ref{shyr}, one concludes
that at least one of $v^fwu$ or $v^gwu$ is primitive, a contradiction.

If $x = y$, then for all integers $n \geq 1$, $v^nwu = x^{ni+j}$ is clearly
non-primitive, and consequently, $uv^nw$ is non-primitive, as required.
Let us now suppose that $uvw$ and $uv^2w$ are $k$-powers for some $k \geq 2$.
Then $vwu = x^{i+j}$ and $v^2wu = x^{2i+j}$ are both $k$-powers as well.
We claim that the following must hold:
\begin{eqnarray*}
i + j & \equiv & 0 \pmod k \\
2i + j & \equiv & 0 \pmod k.
\end{eqnarray*}
To see this, write $vwu = z^k$ for some word $z$.  Then $z^k = x^{i+j}$,
so by Corollary~\ref{ls_cor} $z$ and $x$ are powers of a common word.
Since $x$ is primitive it follows that $z$ is a power of $x$.
In particular, $|x|$ divides $|z|$ and $i + j$ is a multiple of $k$,
as claimed.  A similar argument applies to $v^2wu$.

We conclude that $i \equiv j \equiv 0 \pmod k$,
and hence, $v = x^i$ is a $k$-power.  Moreover, $v^nwu = x^{ni+j}$ is also a
$k$-power for all integers $n \geq 1$, and consequently, $uv^nw$ is a
$k$-power, as required.
\end{proof}

The characterization due to Ito et al.
\cite[Proposition~10]{Ito&Katsura&Shyr&Yu:1988} (see also D\"om\"osi,
Horv\'ath, and Ito \cite[Theorem~3]{Domosi&Horvath&Ito:2004})
of the regular languages consisting only of powers,
along with Theorem~\ref{slender}, implies
that any such language is slender.  A simple application of the
Myhill--Nerode Theorem gives the following weaker result.

\begin{proposition}
\label{my-ner}
Let $L$ be a regular language and let $k \geq 2$ be an integer.  If
all but finitely many words of $L$ are $k$-powers, then $L$ is slender.
In particular, if $L$ is accepted by an $n$-state DFA and all words in $L$ of
length $\geq \ell$ are $k$-powers, then for all $r \geq \ell$,
the number of words in $L$ of length $r$ is at most $n$.
\end{proposition}

\begin{proof}
Let $x^k$ and $y^k$ be distinct words in $L$ of length $r \geq \ell$.
Then $x$ and $y$ are inequivalent with respect to the Myhill--Nerode
equivalence relation, since $y^k \in L$ but $xy^{k-1} \not\in L$.
The Myhill--Nerode equivalence relation on $L$ thus has index at least
as large as the number of distinct words of length $r$ in $L$.  Since
the index of the Myhill--Nerode relation is at most $n$, it follows that
there is a bounded number of words of length $r$ in $L$, so that $L$
is slender, as required.
\end{proof}

The following characterization is analogous to the characterization
of palindromic regular languages given in
\cite[Theorem~8]{Horvath&Karhumaki&Kleijn:1987}.

\begin{theorem}
Let $L \subseteq \Sigma^*$ be a regular language and let $k \geq 2$ be
an integer.  The language $L$ consists only of $k$-powers if and only if
it can be written as a finite union of languages of the form
$uv^*w$, where $u,v,w \in \Sigma^*$ satisfy the following:
there exists a primitive word $x \in \Sigma^*$ and integers $i,j \geq 0$
such that $v = x^{ik}$ and $wu = x^{jk}$.
\end{theorem}

\begin{proof}
The ``if'' direction is clear; we prove the ``only if'' direction.
Let $L$ consist only of $k$-powers.  Then by Proposition~\ref{my-ner},
$L$ is slender.  By Theorem~\ref{slender}, $L$ can be written
as a finite union of languages of the form $uv^*w$.  By examining the proof
of Proposition~\ref{prim}, one concludes that $u$, $v$, and $w$ have the
desired properties.
\end{proof}

We shall need the following lemma for the proof of Theorem~\ref{k-pow}.

\begin{lemma}\label{inf_many}
Let $L$ be a regular language accepted by an $n$-state NFA $M$ and let
$k \geq 2$ be an integer.  If $L$ contains a non-$k$-power of length
$\geq n$, then $L$ contains infinitely many non-$k$-powers.
\end{lemma}

\begin{proof}
Let $s \in L$ be a non-$k$-power such that $|s| \geq n$.  Consider
an accepting computation of $M$ on $s$.  Such a computation must contain
at least one repeated state.  It follows that there exists a decomposition
$s = uvw$, $v \neq \epsilon$, such that $uv^*w \subseteq L$.
Let $x$ be the primitive root of $v$, so that $v = x^i$ for some positive
integer $i$.

Suppose that $wu = \epsilon$.  Since $s = v = x^i$ is not a $k$-power,
it follows that $i \not\equiv 0 \pmod k$.  Moreover, there exist
infinitely many positive integers $\ell$ such that
$\ell i \not\equiv 0 \pmod k$, and so by Corollary~\ref{ls_cor}, there
exist infinitely many
words of the form $v^\ell = x^{\ell i}$ that are
non-$k$-powers in $L$, as required.

Suppose then that $wu \neq \epsilon$.
Let $y$ be the primitive root of $wu$, so that $wu = y^j$ for some positive
integer $j$.  We have two cases.

Case 1: $x = y$.  Since $uvw$ is a not a $k$-power, $vwu$ is also not
a $k$-power, and thus we have $i + j \not\equiv 0 \pmod k$.
Moreover, there are infinitely many
positive integers $\ell$ such that $\ell i + j \not\equiv 0 \pmod k$.
For all such $\ell$, the word $v^\ell wu = x^{\ell i + j}$ is not
a $k$-power, and hence the word $uv^\ell w$ is a non-$k$-power in $L$.
We thus have infinitely many non-$k$-powers in $L$, as required.

Case 2: $x \neq y$.  By Theorem~\ref{p+q+}, $v^*wu$ contains
infinitely many primitive words.  Thus, $uv^*w$ contains infinitely
many non-$k$-powers, as required.
\end{proof}

We are now ready to prove Theorem~\ref{k-pow}.

\begin{proof}[Proof of Theorem~\ref{k-pow}]
The proof is similar to that of \cite[Proposition~7]{Ito&Katsura&Shyr&Yu:1988}.
It suffices to prove statement (2) of the theorem, since statement (1)
follows immediately from (2) and Lemma~\ref{inf_many}.

Suppose that $L$ contains infinitely many non-$k$-powers.  Then
$L$ contains a non-$k$-power $s$ with $|s| \geq n$.  Suppose, contrary to
statement (2), that a shortest such $s$ has $|s| > 3n$.
Then any computation of $M$ on
$s$ must repeat some state at least $4$ times.  It follows that
there exists a decomposition $s = u v_1 v_2 v_3 w$,
$v_1,v_2,v_3 \neq \epsilon$, such that $u v_1^* v_2^* v_3^* w \subseteq L$.
We may assume further that $|v_1v_2v_3| \leq 3n$, so that $wu \neq \epsilon$.

Let $p_1$, $p_2$, $p_3$, and $q$ be the primitive roots of
$v_1$, $v_2$, $v_3$, and $wu$, respectively.
Let $v_1 = p_1^{i_1}$, $v_2 = p_2^{i_2}$, $v_3 = p_3^{i_3}$, and $wu = q^j$,
for some integers $i_1,i_2,i_3,j>0$.  We consider three cases.

Case~1: $p_1 = p_2 = p_3 = q$.
Without loss of generality, suppose that $|v_1| \leq |v_2| \leq |v_3|$.
Since $|s| > 3n$, we must have $|uv_3w| \geq n$, and thus
$|uv_1v_3w| \geq n$ and $|uv_2v_3w| \geq n$.  By assumption, the words
$v_3wu = q^{i_3+j}$, $v_1v_3wu = q^{i_1+i_3+j}$, and
$v_2v_3wu = q^{i_2+i_3+j}$ are $k$-powers, whereas the word
$v_1v_2v_3wu = q^{i_1+i_2+i_3+j}$ is not.  Applying Corollary~\ref{ls_cor},
we deduce that the following system of equations
\begin{eqnarray*}
i_1 + i_2 + i_3 + j & \not\equiv & 0 \pmod k \\
i_3 + j & \equiv & 0 \pmod k \\
i_1 + i_3 + j & \equiv & 0 \pmod k \\
i_2 + i_3 + j & \equiv & 0 \pmod k
\end{eqnarray*}
must be satisfied.  However, it is easy to see that this is impossible.

Case~2: $p_1 \neq q$ and $p_2 = p_3 = q$.  If $|v_1wu| \leq n$,
then let $\ell$ be the smallest positive integer
such that $n \leq |v_1^\ell wu| < |v_1^{\ell+1} wu| \leq |s|$.  Then
by Proposition~\ref{shyr}, one of the words $v_1^\ell wu$ or
$v_1^{\ell+1} wu$ is primitive.  Hence,
at least one of the words $u v_1^\ell w$ or $u v_1^{\ell+1} w$
is a primitive word in $L$, contradicting the minimality of $s$.

If, instead, $|v_1wu| > n$, then we have $n < |v_1wu| < |v_1v_2wu| \leq |s|$.
Again, by Proposition~\ref{shyr}, one of the words $v_1wu$ or
$v_1v_2wu$ is primitive.  Hence, at least one of the words
$uv_1w$ or $uv_1v_2w$ is a primitive word in $L$, contradicting
the minimality of $s$.

Case~3: $p_1 \neq q$ and $p_2 \neq q$.  In this case we choose the
smaller of $v_1$ and $v_2$ to ``pump'', so without loss of generality,
suppose $|v_1| \leq |v_2|$.  Let $\ell$ be the smallest positive integer
such that $n \leq |v_1^\ell wu| < |v_1^{\ell+1} wu| \leq |s|$.  Note
that $|v_1^2 wu| \leq |v_1v_2wu| < |s|$, so such an $\ell$ must exist.
Then by Proposition~\ref{shyr}, one of the words $v_1^\ell wu$ or
$v_1^{\ell+1} wu$ is primitive.  Hence,
at least one of the words $u v_1^\ell w$ or $u v_1^{\ell+1} w$
is a primitive word in $L$, contradicting the minimality of $s$.

All remaining possibilities are symmetric to the cases considered above.
Since in all cases we derive a contradiction, it follows that if $L$
contains infinitely many non-$k$-powers, it contains a non-$k$-power
$s$, where $n \leq |s| \leq 3n$.

It remains to consider the situation where $M$ is a DFA over an alphabet
of size $\geq 2$.  Let $a \neq b$ be alphabet symbols of $M$.  If $M$ does
not have a dead state, then for every integer $i \geq n-1$,
there exists a word $x$, $|x| \leq n-1$, such
that $a^ibx \in L$.  These words $a^ibx$ are all distinct and primitive.
Thus, whenever $M$ has no dead state, $M$ always accepts infinitely many
non-$k$-powers, and, in particular, $M$ accepts a non-$k$-power $s$,
where $n \leq |s| \leq 2n-1$.

If, on the other hand, $M$ does have a dead state,
then we may delete this dead state and apply
the earlier argument with the bound $3n-3$ in place of $3n$.

Finally, the converse of statement (2) follows immediately from
Lemma~\ref{inf_many}.
\end{proof}

We can now deduce the following algorithmic result.

\begin{theorem}\label{alg_allkpow}
Let $k \geq 2$ be an integer.  Given an NFA $M$ with $n$ states and
$t$ transitions, it is
possible to determine if every word in $L(M)$ is a $k$-power in
$O(n^3 + t n^2)$ time.
\end{theorem}

\begin{proof}
     The proof is exactly analogous to that of Theorem~\ref{thm2}, and
we only indicate what needs to be changed.  Suppose $M$ has $t$ states.
We create an NFA, $M'_r$, for $r = 3t$, such that
no word in $L(M'_r)$ is a $k$-power, and $M'_r$ accepts all non-$k$-powers
of length $\leq r$ (and perhaps some other non-$k$-powers).

Note that we may assume that $k \leq r$.  If $k > r$, then no word of
length $\leq r$ is a $k$-power.  In this case, to obtain the desired
answer it suffices to test if the set $\{ x \in L(M) : |x| \leq r \}$
is empty.  However, this set is empty if and only if $L(M)$ is empty, and
this is easily verified in linear time.

We now form a new NFA $A$ as
the cross product of $M'_r$ with $M$.  From Theorem~\ref{k-pow}, it follows
that $L(A) = \emptyset$ iff
every word in $L(M)$ is a $k$-power.  We can determine if
$L(A) = \emptyset$ by
checking (using depth-first search)
whether any final states of $A$ are reachable from the start state.

     It remains to see how $M'_r$ is constructed.
If the length of a word $x$ accepted by $M_r'$ is a multiple of $k$, 
$x$ can be partitioned into $k$ sections of equal length.  In order 
for $M_r'$ to accept $x$, the NFA must `verify' a symbol mismatch between 
two symbols found in different sections but in the same position.

If $x$ is a non-$k$-power, then a symbol mismatch will occur between two 
sections of $x$, call them $s_i$ and $s_j$.  This means that $s_i$ and 
$s_j$ differ in at least one position.  Comparing $s_i$ and $s_j$ to 
$s_1$, the first section of $x$, we notice that at least one of $s_i$ or 
$s_j$ must have a symbol mismatch with $s_1$ (otherwise $s_1=s_i=s_j$, 
which would give a contradiction).  Therefore, when checking $x$ for a 
symbol mismatch, it is sufficient to only check $s_1$ against each of the 
remaining $k-1$ sections, as opposed to checking all $k \choose 
2$ possibilities.

In order to construct $M_r'$, we create a series of `lobes', each of which 
is connected to the start state by an $\epsilon$-transition.  Each lobe 
represents three simultaneous `guesses' made by the NFA, which are:

\begin{itemize}
\item Which alphabet symbols will conflict and in which order.  The number 
of possible conflict pairs is $| \Sigma | \left( |\Sigma| - 1 \right)$.

\item The section in which there will be a symbol mismatch with the first 
section.  There are $k-1$ possible sections.

\item The position in which the conflict will occur.  In the worst case 
when the length of the input is $r$, there will be at most
$r/k$ possible positions.
\end{itemize}

This gives a total of at most $|\Sigma|\left( |\Sigma| - 1 \right) \cdot (k-1) 
\cdot  r/k $ lobes.  The construction of each lobe is
illustrated in Figure~\ref{fig:module}.

\begin{figure}[hbt]
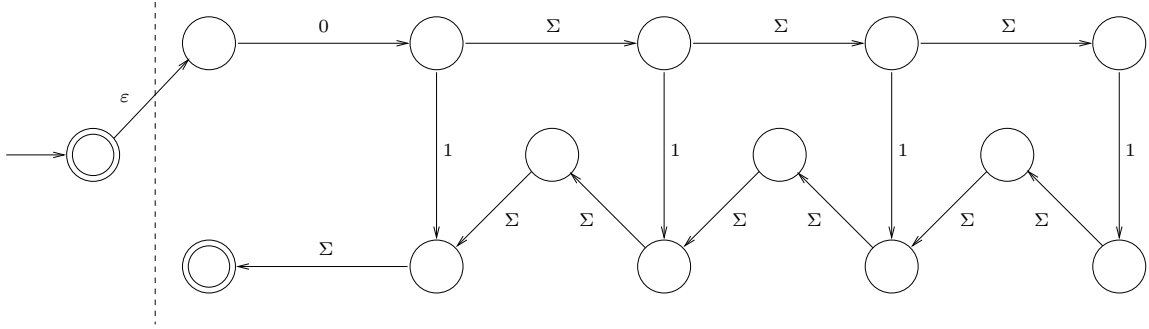

\input module.tex
\caption{One lobe of the NFA for $k=3$, $r=12$ and
$0,1$ conflicting symbols.}
\label{fig:module}
\end{figure}

Each lobe contains at most $r+1$ states.
In addition to these lobes, we also require a $k$-state submachine to
accept all words whose lengths are not a multiple of $k$.

In total, $M_r'$ has at most
$$|\Sigma| \left( |\Sigma| - 1 \right) \cdot (k-1) \cdot 
{r \over k}  \cdot (r+1) + k + 1 \in O(r^2)$$ states
(since $k \leq r$), and similarly, $O(r^2)$ 
transitions.
After constructing the cross-product, this gives a $O(n^3 + tn^2)$ 
bound on the time required to determine if every word in $L(M)$ is a 
$k$-power.
\end{proof}

     Theorem~\ref{k-pow} suggests the following question:  if $M$ is an
NFA with $n$ states that accepts at least one non-$k$-power, how long
can a shortest non-$k$-power be?   Theorem~\ref{k-pow} proves an
upper bound of $3n$.   A lower bound of $2n-1$ for infinitely many
$n$ follows easily from the obvious $(n+1)$-state NFA accepting 
${\tt a}^{n} ({\tt a}^{n+1})^*$, where $n$ is divisible by $k$.  
However, Ito, Katsura, Shyr, and Yu \cite{Ito&Katsura&Shyr&Yu:1988}
gave a very interesting example that improves this lower bound:
if $x = ( (ab)^n a)^2$ and $y = ba x ab$,  then $x$ and $xyx$ are
squares, but $xyxyx$ is not a power.  Hence, the obvious $(8n+8)$-state NFA
that accepts $x(yx)^*$ has the property that the shortest non-$k$-power
accepted is of length $20n+18$.  This improves the lower bound  for
infinitely many $n$.

       We now generalize their lower bound.

\begin{proposition}
       Let $k \geq 2$ be fixed.  There exist infinitely many NFAs $M$
with the property that if $M$ has $r$ states, then the shortest
non-$k$-power accepted is of length $\left(2+ {1 \over{2k-2}}\right) r - O(1)$.
\end{proposition}

\begin{proof}
Let $u = (ab)^n a$, $x = u^k$, and $ y = x^{-1} (x ba u^{-1} x)^k x^{-1}$.
Thus $xyx = (x ba u^{-1} x)^k$.
Hence $x$ and $xyx$ are both $k$-powers.

However, $xyxyx$ is not a $k$-power.  To see this,
assume it is, and write $xyxyx = g_1 g_2 \cdots g_k$.
Look at the character in position $2kn-2n+k$ (indexing beginning with 1)
in $g_1$ and $g_k$.  In $g_1$ it is $a$, and in $g_k$ it is $b$, so
$xyxyx$ is not a $k$-power.

We can accept $x(yx)^*$ with an NFA using $|xy|$ states.  
The shortest non-$k$-power is $xyxyx$, which is of length $m$.

We have $|u| = 2n+1$, $|x| = k(2n+1)$, $|y| = k(4kn - 6n + 2k - 1)$,
$ r = |xy| = 2k(2kn - 2n + k)$, and $ m = |xyxyx| = k(8kn - 6n + 4k + 1)$.
Thus $m = {{4k-3} \over {2k-2}}r - {k \over {k-1}} =
\left(2 + {1 \over {2k-2}}\right)r - O(1)$.
\end{proof}

Next, we apply part~(2) of Theorem~\ref{k-pow} to obtain an algorithm
to check if an NFA accepts infinitely many non-$k$-powers.

\begin{theorem}
Let $k \geq 2$ be an integer.  Given an NFA $M$ with $n$ states and
$t$ transitions, it is possible to determine if all but finitely many words
in $L(M)$ are $k$-powers in $O(n^3 + t n^2)$ time.
\end{theorem}

\begin{proof}
The proof is similar to that of Theorem~\ref{alg_allkpow}.
The only difference is that in view of part~(2) of Theorem~\ref{k-pow}
we instead construct $M_r'$ to accept all non-$k$-powers $s$,
where $n \leq |s| \leq 3n$.  We leave the details to the reader.
\end{proof}

\section{Automata accepting only powers}
\label{powers}

In this section we move from the problem of testing if an automaton
accepts only $k$-powers to the problem of testing if it accepts only
powers (of any kind).  Just as Theorem~\ref{k-pow} was the starting
point for our algorithmic results in Section~\ref{kp}, the
following theorem of Ito, Katsura, Shyr, and Yu
\cite{Ito&Katsura&Shyr&Yu:1988} is the starting point for our
algorithmic results in this section.  We state the theorem in
a stronger form than was originally presented by Ito et al.

\begin{theorem}
\label{ito}
Let $L$ be accepted by an $n$-state NFA $M$.
\begin{enumerate}
\item Every word in $L$ is a power if and only if every word in the set
$\lbrace x \in L : |x| \leq 3n \rbrace$ is a power.
\item All but finitely many words in $L$ are powers if and only if
every word in the set $\lbrace x \in L : n \leq |x| \leq 3n \rbrace$
is a power.
\end{enumerate}
Further, if $M$ is a DFA over an alphabet of size $\geq 2$, then the bound $3n$
may be replaced by $3n-3$.
\end{theorem}

We next prove an analogue of Proposition~\ref{my-ner}.  We
need the following result, first proved by Birget \cite{Birget:1992},
and later, independently, in a weaker form, by Glaister and Shallit
\cite{Glaister&Shallit:1996}.

\begin{theorem}
\label{birget}
Let $L \subseteq \Sigma^*$ be a regular language.  Suppose there exists
a set of pairs
\[
S = \{(x_i,y_i) \in \Sigma^* \times \Sigma^* : 1 \leq i \leq n \}
\]
such that
\begin{itemize}
\item $x_iy_i \in L$ for $1 \leq i \leq n$, and
\item either $x_iy_j \notin L$ or $x_jy_i \notin L$ for $1 \leq i,j \leq n$,
$i \neq j$.
\end{itemize}
Then any NFA accepting $L$ has at least $n$ states.
\end{theorem}

\begin{proposition}
\label{slender_7n}
Let $M$ be an $n$-state NFA and let $\ell$ be a non-negative integer
such that every word in $L(M)$ of length $\geq \ell$ is a power.
For all $r \geq \ell$, the number of words in $L(M)$ of length $r$
is at most $7n$.
\end{proposition}

\begin{proof}
Let $r \geq \ell$ be an arbitrary integer.  The proof consists of three steps.

Step~1.  We consider the set $A$ of words $w$ in $L(M)$ such that
$|w| = r$ and $w$ is a $k$-power for some $k \geq 4$.  For each such $w$,
write $w = x^i$, where $x$ is a primitive word, and define a pair
$(x^2,x^{i-2})$.  Let $S_A$ denote the set of such pairs.
Consider two pairs in $S_A$: $(x^2,x^{i-2})$ and
$(y^2,y^{j-2})$.  The word $x^2y^{j-2}$ is primitive by Theorem~\ref{ls_eqn}
and hence is not in $L(M)$.  The set $S_A$ thus satifies the conditions
of Theorem~\ref{birget}.  Since $L(M)$ is accepted by an $n$-state
NFA, we must have $|S_A| \leq n$ and thus $|A| \leq n$.

Step~2.  Next we consider the set $B$ of cubes of length $r$ in $L(M)$.
For each such cube $w = x^3$, we define a pair $(x,x^2)$.  Let
$S_B$ denote the set of such pairs.  Consider two pairs in $S_B$:
$(x,x^2)$ and $(y,y^2)$.  Suppose that $xy^2$ and $yx^2$ are both in
$L(M)$.  The word $xy^2$ is certainly not a cube; we claim that it
cannot be a square.  Suppose it were.  Then $|x|$ and $|y|$ are even,
so we can write $x = x_1 x_2$ and $y = y_1 y_2$ where
$|x_1| = |x_2| = |y_1| = |y_2|$.  Now if $xy^2 = x_1 x_2 y_1 y_2 y_1 y_2$ 
is a square, then $x_1 x_2 y_1 = y_2 y_1 y_2$, and so $y_1 = y_2$.
Thus $y$ is a square; write $y = z^2$.
By Theorem~\ref{ls_eqn}, $yx^2 = z^2x^2$ is primitive,
contradicting our assumption that $yx^2 \in L(M)$.  It must be
the case then that $xy^2$ is a $k$-power for some $k \geq 4$.
Thus, $xy^2 = u^k$ for some primitive $u$ uniquely determined by $x$ and $y$.
With each pair of cubes $x^3$ and $y^3$ such that both $xy^2$ and $yx^2$
are in $L(M)$ we may therefore associate a $k$-power $u^k \in L(M)$ of length
$r$, where $k \geq 4$.  We have already established in Step~1 that
the number of such $k$-powers is at most $n$.  It follows that
by deleting at most $n$ pairs from the set $S_B$ we obtain
a set of pairs satisfying the conditions of Theorem~\ref{birget}.
We must therefore have $|S_B| \leq 2n$ and thus $|B| \leq 2n$.

Step~3.  Finally we consider the set $C$ of squares of length $r$ in
$L(M)$.  For each such square $w = x^2$, we define a pair $(x,x)$.
Let $S_C$ denote the set of such pairs.  Consider two pairs in
$S_C$: $(x,x)$ and $(y,y)$.  Suppose that $xy$ and $yx$ are both
in $L(M)$.  The word $xy$ is not a square and must therefore be
a $k$-power for some $k \geq 3$.  We write $xy = u^k$ for some
primitive $u$ uniquely determined by $x$ and $y$.  In Steps~1 and 2
we established that the number of $k$-powers of length $r$, $k \geq 3$,
is $|A| + |B| \leq 3n$.  It follows that
by deleting at most $3n$ pairs from the set $S_C$ we obtain
a set of pairs satisfying the conditions of Theorem~\ref{birget}.
We must therefore have $|S_C| \leq 4n$ and thus $|C| \leq 4n$.

Putting everything together, we see that there are
$|A| + |B| + |C| \leq 7n$ words of length $r$ in $L(M)$,
as required.
\end{proof}

     The bound of $7n$ in Proposition~\ref{slender_7n} is almost
certainly not optimal.  

We now prove the following algorithmic result.

\begin{theorem}
Given an NFA $M$ with $n$ states, it is
possible to determine if every word in $L(M)$ is a power in
$O(n^5)$ time.
\label{kats}
\end{theorem}

\begin{proof}
     First, we observe that we can test whether a word $w$ of length
$n$ is a power in $O(n)$ time, using a linear-time string matching
algorithm, such as Knuth-Morris-Pratt \cite{Knuth&Morris&Pratt:1977}.  
To do so, search for $w = a_1 a_2 \cdots a_n$ in the word
$x = a_2 \cdots a_n a_1 \cdots a_{n-1}$.  Then $w$ appears in $x$ iff
$w$ is a power.  Furthermore, if the leftmost occurrence of
$w$ in $x$ appears beginning at $a_i$, then $w$ is a $n/(i-1)$ power, and
this is the largest exponent of a power that $w$ is.

     Now, using Theorem~\ref{ito}, it suffices to test all words
in $L(M)$ of length $\leq 3n$;  every word in $L(M)$ is a power iff all
of these words are powers.  On the other hand, by
Proposition~\ref{slender_7n}, if all words are powers, then
the number of words of each length is bounded by $7n$.  Thus, it
suffices to enumerate the words in $L(M)$ of lengths $1,2, \ldots, 3n$,
stopping if the number of such words in any length exceeds $7n$.  If all
these words are powers, then every word is a power.  Otherwise, if we
find a non-power, or if the number of words in any length exceeds $7n$,
then not every word is a power.

      By the work of M\"akinen \cite{Makinen:1997} or 
Ackerman \& Shallit \cite{Ackerman&Shallit:2007}, we can enumerate 
these words in $O(n^5)$ time.
\end{proof}

      Using part~(2) of Theorem~\ref{ito} along with
Proposition~\ref{slender_7n}, we can prove the following.

\begin{theorem}
     Given an NFA $M$ with $n$ states,
we can decide if all but finitely many words in $L(M)$ are
non-powers in $O(n^5)$ time.
\end{theorem}

\begin{proof}
      The proof is analogous to that of Theorem~\ref{kats}.  The only
difference is that here we need only enumerate the words in $L(M)$ of
lengths $n,n+1,\ldots,3n$.
\end{proof}

\section{Bounding the length of a smallest power}
\label{smallkp}

In Section~\ref{kp} we gave an upper bound on the length of
a smallest non-$k$-power accepted by an $n$ state NFA.  In this section
we study the complementary problem of bounding the length of
the smallest $k$-power accepted by an $n$-state NFA.

\begin{proposition}
\label{upper_bd}
Let $M$ be an NFA with $n$ states and let $k \geq 2$ be an integer.
If $L(M)$ contains a $k$-power, then $L(M)$ contains a $k$-power
of length $\leq kn^k$.
\end{proposition}

\begin{proof}
Consider the NFA-$\epsilon$ $M'$ accepting $L(M)^{1/k}$ defined in the proof of
Proposition~\ref{fixed-k}.  The only transitions from the start
state of $M'$ are $\epsilon$-transitions to submachines whose states are
$(2k-1)$-tuples of the form
$[g_1, g_2, \ldots, g_{k-1}, p_0, p_1, \ldots, p_{k-1}]$,
where the first $(k-1)$-elements of the tuple are fixed.  Thus we may
consider $L(M')$ as a finite union of languages, each accepted by
an NFA of size $n^k$.  It follows that if $M'$ accepts a non-empty
word $w$, it accepts such a $w$ of length $\leq n^k$.  However,
$M'$ accepts $w$ if and only if $M$ accepts $w^k$.  We conclude that
if $L(M)$ contains a $k$-power, it contains one of length $\leq kn^k$.
\end{proof}

We now give a lower bound on the size of the smallest $k$-power
accepted by an $n$-state DFA.

\begin{proposition}
Let $k \geq 2$ be an integer.  There exist infinitely many DFAs
$M_n$ such that

\begin{itemize}
\item[(a)] $M_n$ has $O(kn)$ states;
\item[(b)] The shortest $k$-power accepted by $M_n$ is of length
$k\cdot\Omega\left({n \choose k}\right)$.
\end{itemize}
\end{proposition}

\begin{proof}
For $n \geq k$, let
\[
L_n = ({\tt a}^n)^+ {\tt b} ({\tt a}^{n-1})^+ {\tt b} \cdots
({\tt a}^{n-k+1})^+ {\tt b}.
\]
Then $L_n$ is accepted by a DFA with $O(kn)$ states,
and the shortest $k$-power in $L_n$ is $({\tt a}^\ell{\tt b})^k$,
where
\[
\ell = \text{lcm}(n,n-1,\ldots,n-k+1) \geq n(n-1)\cdots(n-k+1)/k!
= {n \choose k},
\]
as required.
\end{proof}

Next we consider the length of a smallest power (rather than $k$-power).

\begin{proposition}
\label{exponent_bd}
Let $M$ be an NFA with $n$ states.  If $L(M)$ contains a power,
it contains a $k$-power for some $k$, $2 \leq k \leq n+1$.
\end{proposition}

\begin{proof}
Suppose to the contrary
that the smallest $k$ for which $L(M)$ contains a $k$-power $w^k$
satisfies $k > n+1$.  For some accepting computation of $M$ on $w^k$ let
$q_1,q_2,\ldots,q_{k-1}$ be the states reached by $M$ after
reading $w,w^2,\ldots,w^{k-1}$ respectively.  Since $k > n+1$, there
exist $i$ and $j$ where $1 \leq i < j \leq k-1$ and $q_i = q_j$.
It follows that $M$ accepts $w^\ell$ for some $\ell$, $2 \leq \ell < k$,
contradicting the minimality of $k$.  We conclude that if $L(M)$ contains a
$k$-power, we may take $k \leq n+1$.
\end{proof}

\begin{proposition}
Let $M$ be an NFA with $n$ states.  If $L(M)$ contains a power,
then $L(M)$ contains a power of length $\leq (n+1)n^{n+1}$.
\end{proposition}

\begin{proof}
Apply Propositions~\ref{exponent_bd} and \ref{upper_bd}.
\end{proof}

We now give a lower bound.

\begin{proposition}
\label{smallest_pow}
There exist infinitely many DFAs $M_n$ such that

\begin{itemize}
\item $M_n$ has $O(n)$ states;
\item The shortest power accepted by $M_n$ is of length
$e^{\Omega(\sqrt{n \log n})}$.
\end{itemize}
\end{proposition}

\begin{proof}
Let $p_i$ denote the $i$-th prime number.  For any integer
$n \geq 2$, let $P(n) = p_k$ be the largest prime number such that
$p_1 + p_2 + \cdots + p_k \leq n$.  We define
\[
L_n = ({\tt a}^{p_1})^+ {\tt b} ({\tt a}^{p_2})^+ {\tt b} \cdots
({\tt a}^{p_k})^+ {\tt b}.
\]
Then $L_n$ is accepted by a DFA with $O(n)$ states.

If $k$ is itself prime,
the shortest power in $L_n$ is $w = ({\tt a}^\ell{\tt b})^k$,
where $\ell = p_1p_2 \cdots p_k$.  For $n \geq 2$, let
\[
F(n) = \prod_{p \leq P(n)} p,
\]
where the product is over primes $p$.
We have $F(n) \in e^{\Omega(\sqrt{n \log n})}$ \cite[Theorem~1]{Miller:1987}.
This lower bound is valid
for all sufficiently large $n$; in particular, it holds for infinitely
many $n$ such that $n = p_1 + p_2 + \cdots + p_k$, where $k$ is prime.
This gives the desired result.
\end{proof}

\section{Additional results on powers}
\label{add2pow}

D\"om\"osi, Mart\'{\i}n-Vide, and Mitrana
\cite[Theorem~10]{Domosi&Martin-Vide&Mitrana:2004} proved that if $L$
is a slender regular language over $\Sigma$, and $Q_\Sigma$ is the
set of primitive words over $\Sigma$, then $L \cap Q_\Sigma$ is regular.
This result is somewhat surprising, since it is widely believed
that $Q_\Sigma$ is not even context-free for $|\Sigma| \geq 2$.  In this
section we apply a variation of their argument to show that $Q_\Sigma$ may be
replaced by the language of squares, (cubes, etc.) over $\Sigma$.

For any integer $k \geq 2$ and alphabet $\Sigma$, let $P(k,\Sigma)$
denote the set of $k$-powers over $\Sigma$.  Clearly, for $|\Sigma| \geq 2$,
$P(k,\Sigma)$ is not context-free.

\begin{proposition}
If $L \subseteq \Sigma^*$ is a slender regular language, then for all
integers $k \geq 2$, $L \cap P(k,\Sigma)$ is regular.
\end{proposition}

\begin{proof}
If $L$ is slender, then by Theorem~\ref{slender} it
suffices to consider $L = uv^*w$.
The result is clearly true if $v$ is empty, so we suppose $v$ is
non-empty.  Let $x$ and $y$ be the primitive roots of $v$ and $wu$
respectively.  If $x = y$, then the set of $k$-powers in $v^*wu$
is given by $v^*wu \cap (x^k)^*$, so the set of $k$-powers in $uv^*w$
is regular.  If $x \neq y$, then by Theorem~\ref{p+q+},
the set $v^*wu$ contains only finitely many $k$-powers.
The set of $k$-powers in $uv^*w$ is therefore finite, and,
a fortiori, regular.
\end{proof}

\section{Testing if an NFA accepts a bordered word}
\label{bord}

In this section we give an efficient algorithm to test if an
NFA accepts a bordered word.  We also give upper and lower
bounds on the length of a shortest bordered word accepted by
an NFA.

\begin{proposition}
      Given an NFA $M$ with $n$ states and $t$ transitions,
we can decide if $M$ accepts at least one
bordered word in $O(n^3 t^2)$ time.
\label{border}
\end{proposition}

\begin{proof}
      Given an NFA $M = (Q, \Sigma, \delta, q_0, F)$,
      we can easily create an NFA-$\epsilon$ $M'$ that
accepts 
$$\lbrace u \in \Sigma^* \ : \ \text{there exists
$w \in \Sigma^*$ such that } uwu \in L \rbrace$$
by ``guessing'' the state we would be in after reading $uw$, and
then verifying it.   More formally, we let $M' = (Q', \Sigma, 
\delta', q'_0, F')$ where $Q' = \lbrace q'_0 \rbrace
\cup \ \lbrace [p,q,r] \ : \ p, q, r \in Q \rbrace$,
$F' = \lbrace [p,q,r] \ : \ r \in F \text{ and there exists
$w \in \Sigma^*$ such that } q \in \delta(p,w) \rbrace$.
The transitions are defined as follows:
$\delta(q'_0, \epsilon) = \lbrace [q_0, p, p] \ : \ p \in Q \rbrace$
and
$$\delta([p,q,r],a) = \lbrace [p', q, r'] \ : \ p' \in \delta(p,a),
	r' \in \delta(r,a) \rbrace.$$
If $M$ has $n$ states and $t$ transitions,
then $M'$ has $n^3 + 1$ states and at most $n + n^3 t^2$ transitions.
Now get rid of all useless states and their associated transitions.
We can compute the final states by doing $n$ depth-first searches,
starting at each node, at a cost of $O(n(n+t))$ time.  
Now we just test to see if $L(M')$ accepts a nonempty
string, which can be
done in linear time in the size of $M'$.
\end{proof}

\begin{corollary}
     If $M$ is an NFA with $n$ states, and it accepts at least one
bordered word, it must accept a bordered word of length
$< 2n^2 + n$.
\end{corollary}

\begin{proof}
    Consider the DFA $M'$ constructed in the proof of the
previous theorem, which accepts
$$L' = \lbrace u \in \Sigma^* \ : \ \text{there exists
$w \in \Sigma^*$ such that } uwu \in L \rbrace.$$
If $M$ accepts a bordered string, then $M'$ accepts a nonempty string.
Although $M'$ has $n^3+1$ states, once a computation 
leaves $q'_0$ and enters a triple of the form $[p,q,r]$, it never
enters a state $[p',q',r']$ with $q \not= q'$.  Thus we may view
the NFA $M'$ as implicitly defining a union of $n$ disjoint languages,
each accepted by an NFA with $n^2$ states.     Therefore, if $M'$
accepts a nonempty string $u$, it accepts one of length at most $n^2$.
Now the corresponding bordered string is $uwu$.  The string $w$
is implicitly defined in the previous proof as a path from a state
$p$ to a state $q$.  If such a path exists, it is of length at most
$n-1$.  Thus there exists $uwu \in L(M)$  with $|uwu| \leq 2n^2 + n-1$.
\end{proof}

\begin{proposition}
       For infinitely many $n$ there is an DFA of $n$ states
such that the shortest bordered word accepted is of length
$n^2/2 - 6n +43/2$.  
\end{proposition}

\begin{proof}
Consider $a (b^t)^+ c a (b^{t-1})^+ c$.  An obvious DFA can accept
this using $2t+5$ states.  However, the 
shortest bordered word accepted is $a b^{t(t-1)} c 
a b^{t(t-1)} c$, which is of length $2t(t-1)+ 4 = n^2/2 - 6n + 43/2$.
\end{proof}

    We now consider
testing if an NFA accepts infinitely many bordered words.  

\begin{corollary}
     If an NFA $M$ has $n$ states and $t$ transitions,
we can test whether $M$ accepts infinitely many bordered words
in $O(n^6 t^2)$ time.
\end{corollary}

\begin{proof}
      If an NFA $M$ accepts infinitely many words of the form $uwu$,
there are two possibilities, at least one of which must hold:

\begin{itemize}
\item[(a)] there is a single word $u$ such
that there are infinitely many $w$ with $uwu \in L(M)$, or

\item[(b)] there
are infinitely many $u$, with possibly different $w$ depending on $u$,
such that $uwu \in L(M)$.  
\end{itemize}

      To check these possibilities, we return to the NFA-$\epsilon$ $M'$
constructed in the proof of Theorem~\ref{border}.  First, for each pair
of states $q_i$ to $q_j$, we determine whether there exists a nonempty
path from $q_i$ to $q_j$.  This can be done with
$n$ different depth-first searches, starting at each vertex, at a cost
of $O(n^3(n^3+t^2))$ time.
In particular, for each vertex, we learn whether there
is a nonempty cycle beginning and ending at that vertex.

Now let us check whether (a) holds.  After removing all useless states
and their associated transitions, look at the remaining final states
$[p,q,r]$ of $M'$ and determine if there is a path from $p$ to $q$
that goes through a vertex with a cycle.   This can be done by
testing, for each vertex $s$ that has a cycle, whether there is a non-empty
path from $p$ to $s$ and then $s$ to $q$.  If such a vertex exists, then
there are infinitely many $w$ in some $uwu$.  

To check whether (b) holds, we just need to know whether $M'$ accepts
infinitely many strings, which we can easily check by looking for a 
directed cycle.

The total cost is therefore $O(n^3(n^3 t^2))$.
\end{proof}

We now prove the following decomposition theorem for regular languages
consisting only of bordered words.

\begin{theorem}
If every word in a regular language $L$ is bordered, then there is a 
decomposition of $L$ as a finite union of regular languages of the
form $JKJ$, where each $J$ and $K$ are regular and $\epsilon \not\in J$.
\end{theorem}

\begin{proof}
Let $L$ be accepted by an NFA $M = (Q,\Sigma,\delta,q_0,F)$.
For each $x \in \Sigma^+$, define an automaton $M_x = (Q,\Sigma,\delta,I',F')$
(for $M_x$ we permit multiple initial states), where the set of
initial states is $I' = \delta(q_0, x)$,
and the set of final states is $F' = \{q \in Q : \delta(q,x) \in F\}$.
Then $M_x$ has the property that for every $w \in L(M_x)$, we have
$xwx \in L(M)$.  Note that there are only finitely many distinct automata
$M_x$.  

For each automaton $M_x$, define the regular language
\[
L_x = \{y : \delta(q,y) = I' \text{ and } \{q \in Q: \delta(q,y) \in F\} = F'\}.
\]
Note that again there are only finitely many distinct languages $L_x$.

For every $x \in \Sigma^+$, every word in $L_x L(M_x) L_x$ is in $L$.
Furthermore, if $w \in L$ is bordered, then there exists $x \in \Sigma^+$
such that $w \in L_x L(M_x) L_x$.  Thus, if every word of $L$
is bordered, then $L = \cup_{x \in \Sigma^+} L_x L(M_x) L_x$.
Since there are only finitely many languages $L_x$ and $L(M_x)$,
this union is finite, as required.
\end{proof}

\section{Testing if an NFA accepts an unbordered word}
\label{unbord}

We present a simple test to determine if all words in a regular language 
are bordered, and to determine if a regular language contains infinitely many 
unbordered words.     
We first need the following well-known result about words, which is due to 
Lyndon and Sch\"utzenberger \cite{Lyndon&Schutzenberger:1962}.

\begin{lemma}\label{loft1}
Suppose $x$, $y$ and $z$ are non-empty words, and that $xy = yz$.  Then 
there is a non-empty word $p$, a word $q$ and a non-negative 
integer $k_1$ for which we can write $x = pq$, $z = qp$, and $y = (pq)^{k_1}p$.
\end{lemma}
  
We also need the following result, which is just a variation of the 
pumping lemma. 

\begin{lemma}\label{loft2}
Let $M = (Q,\Sigma,\delta,q_0,F)$ be an $n$-state NFA.
Let $L$ be the language accepted by $M$.
Let $d$ be a positive integer. 
Let $(X,y,Z)$ be a $3$-tuple of words 
for which $|y|$ is a multiple of $d$, $|y| \ge nd$ and $XyZ \in L$.  
Then there are words $r$, $s$ and $t$, whose lengths are multiples of $d$,
with $|s| \ge d$, for which we can 
write $y = rst$, and, for all $z \ge 0$, $Xrs^ztY \in L$.    
\end{lemma}

\begin{proof}
Set $l := |X|$ and $m := |y|/d$, $\gamma := XyZ$, and $k := |\gamma|$.    
First, write $\gamma$ as a sequence of letters, that is, 
$\gamma := \gamma_1 \gamma_2 \cdots
\gamma_k$ with each $\gamma_i$ a letter.  By $\gamma[i,j]$ for $1 \le i,j 
\le |\gamma|$ we
mean the subsequence that consists of the $i-j+1$ consecutive letters of $\gamma$ 
starting at position $i$ and ending at position $j$, that is, $\gamma_i 
\gamma_{i+1}\cdots \gamma_j$.
If $i > j$ we take $\gamma[i,j]$ to be the empty word.
Now we have the following sequence of $k$ states
\[q_1 \in \delta(q_0, \gamma_1), q_2 \in \delta(q_1, \gamma_2), \dots,
q_k \in \delta(q_{k-1}, \gamma_k).\]
We'll choose $q_k$ to be a final state.

Note that $y = \gamma[l+1, l+md]$, and consider the following sequence 
of $m+1$ states of $M$:

\[q_l, q_{l+d}, q_{l+2d}, \dots, q_{l+md}.\]

There are integers $i$ and $j$, with $0 \le i < j \le m$ for which 
$q_{l+id} = q_{l+jd}$.  Set $r := \gamma[l+1, l+id]$, $s := \gamma[l+id+1, 
l+jd]$, and $t := \gamma[l+jd+1, l+md]$, so $y = rst$.  Note that $|s| \ge d$, 
and the desired conclusion follows immediately.    
\end{proof}

\begin{lemma}\label{loft3}
Let $M$ be an $n$-state NFA.  Let $L$ be the language accepted by $M$.
Let $(X,Y,Z)$ be a $3$-tuple of words for which $XYZ \in L$.
Then there is a word $y$ for which $|y| < n$ and $XyZ \in L$.  
\end{lemma}
    
\begin{proof}
Let $S := \{u \in \Sigma^{*} : XuZ \in L \}$.  Let $y$ be an element of 
$S$ of minimal length.  We proceed by contradiction, and suppose $|y| \ge n$.
We apply Lemma~\ref{loft2} to $(X,y,Z)$, with $d = 1$, and write $y = rst$
with $s$ non-empty.  Then $XrtZ \in L$, which violates the minimality of $|y|$.
\end{proof}

\begin{lemma}\label{loft4}
Suppose there are words $\Psi_L$, $\Psi_R$, $e$, $f$, $g$ and $h$ with
$|\Psi_L| = |\Psi_R|$, $|e| < |\Psi_L|$, $|g| < |\Psi_L|$, and for which 
\begin{equation}\label{star1}
b_\zeta := \Psi_Le = f\Psi_R,
\end{equation}
and
\begin{equation}\label{star2}
b_\eta := \Psi_Lg = h\Psi_R.  
\end{equation}
Suppose further that $|b_\eta| < |b_\zeta|$.  
Then we can write $\Psi_L = h(pq)^{k}p$ and $\Psi_R = (pq)^{k}pg$ 
for $p$ a non-empty word, $q$ a word for which $|g| + |pq| = |f|$,
and $k$ a positive integer. 
\end{lemma}    

\begin{proof}
Since $|b_\eta| < |b_\zeta|$, we must have $|g| < |e| < |\Psi_R|$.  
This last observation, together with (\ref{star1}) and (\ref{star2}) 
above allows us to assert that there are non-empty words $s_1$ and $s_2$, with 
$|s_2| > |s_1|$, such that $\Psi_R = s_1e = s_2g$.
This last fact combined again with (\ref{star1}) and (\ref{star2}) yields that
\begin{equation}\label{star3}
\Psi_L = f s_1 = hs_2,
\end{equation}
and
\begin{equation}\label{star4}
\Psi_R = s_1e  = s_2g.
\end{equation}

Now we can apply (\ref{star3}) and (\ref{star4}) to assert that there are
non-empty words $r_1$ and $r_2$ for which $s_1 r_1 = s_2 = r_2 s_1$; that is,
\begin{equation}\label{star5}
s_1 r_1 = r_2 s_1.  
\end{equation}

Now apply Lemma~\ref{loft1} to (\ref{star5}) to get that there is a non-empty
word $p$, a word $q$ and an integer $k_1 \ge 0$ for which
$s_1 = (pq)^{k_1}p$, $r_1 = qp$, and $r_2 = pq$.  Set $k := k_1 + 1$.
Then $s_2 = (pq)^{k}p$, and (\ref{star3}) gives $\Psi_L = h(pq)^{k}p$,
and (\ref{star4}) gives $\Psi_R = (pq)^{k}pg$.  
Also $s_2 = r_2 s_1$ combined with (\ref{star3}) above gives that $f = hr_2$, 
so $|g| + |pq| = |h| + |pq| = |h| + |r_2| = |f|$. 
\end{proof}

Theorems~\ref{loft_thm1} and \ref{loft_thm3} below are the main results. 

\begin{theorem}\label{loft_thm1}
Let $M$ be an $n$-state NFA. Let $L$ be the language accepted by $M$.  
Let $N$ be a non-negative integer.  
Suppose all words in $L$ of length in the interval $[N, 2N+6n+1]$ are bordered.
Then all words in $L$ of length greater than $2N+6n+1$ are bordered. 
Hence, if all words in $L$ of length at most $6n+1$ are bordered, then all the words 
in $L$ must be bordered. 
\end{theorem}

\begin{proof}
We'll prove Theorem~\ref{loft_thm1} by making  the following series
of observations.
Throughout, we'll assume that all words in $L$ of length in the interval 
$[N, 2N + 6n+1]$ are bordered, and we'll assume $w$ is an unbordered word in $L$
for which $|w| > 2N+6n+1$, with $|w|$ minimal.  We write $w$ as $u \theta v$ with 
$\theta$ a word for which $|\theta| \le 1$ and $u$ and $v$ words for 
which $|u| = |v| > 3n + N$. 

\begin{claim}\label{claim1}
Write $u$ as $\Psi_L X_L$ and $v$ as
$X_R \Psi_R$, for words $\Psi_L$, $X_L$, 
$\Psi_R$, $X_R$ for which $|X_L| = |X_R| = n$.  
(So that $w$ is  $\Psi_L X_L \theta X_R \Psi_R$.)
Then there are words $x_L$ and $x_R$, both of length less than $n$, for 
which:
\begin{itemize}
\item[(i)] $\zeta := \Psi_L x_L \theta X_R \Psi_R \in L$, and
\item[(ii)] $\eta := \Psi_L X_L \theta x_R \Psi_R \in L$.
\end{itemize}
Further, $N \le |\zeta | < |w|$, and $N \le |\eta| < |w|$.  
\end{claim}

To justify (i), apply Lemma~\ref{loft3} to the 3-tuple $(\Psi_L, X_L, 
\theta X_R \Psi_R)$.  Similarly, to arrive at (ii), apply Lemma~\ref{loft3}
again to the 3-tuple $(\Psi_L X_L \theta, X_R, \Psi_R)$.     

\begin{claim}\label{claim2}
We can write $\Psi_L = h(pq)^{k}p$ and $\Psi_R = (pq)^{k}pg$ 
for $p$ a non-empty word, $g$, $h$ and $q$ words for which $|g| = |h|$, 
$|pq| + |g| \le n$, and $k$ a positive integer. 
Hence $w$ can be written as $h(pq)^{k}p X_L \theta X_R (pq)^{k}pg$. 
\end{claim}

To justify Claim~\ref{claim2}, first recall $w = \Psi_L X_L \theta X_R 
\Psi_R$ and $|\Psi_L| = |\Psi_R| > 2n$. 
From Claim~\ref{claim1} above we get that $\zeta$ and $\eta$
are bordered words, so we can assert that there exist non-empty words
$b_{\zeta}$ and  $b_{\eta}$, and words $p_{\zeta}$ and $p_{\eta}$, for which:
\begin{itemize}
\item[(I)]    $\zeta = \Psi_L x_L \theta X_R 
\Psi_R = b_{\zeta} p_{\zeta} b_{\zeta}$, and 
\item[(II)]   $\eta = \Psi_L X_L \theta x_R \Psi_R = 
b_{\eta}p_{\eta} b_{\eta}$.
\end{itemize}

Note that, if  $|b_{\zeta}| \le |\Psi_L|$ then by (I) $b_{\zeta}$ 
would be a border for $w$.  So we must have  $|b_{\zeta}| > |\Psi_L|$.
Similarly, (II) gives that  $|b_{\eta}| > |\Psi_L|$.
These latter facts together with (I) and (II) give that there exists 
non-empty words $e$, $f$, $g$, $h$, for which $|e| = |f|$, $|g| = |h|$,
and for which
\begin{equation}\label{2nd_star1}
b_{\zeta} = \Psi_L e = f \Psi_R,
\end{equation}
and
\begin{equation}\label{2nd_star2}
b_{\eta} = \Psi_L g = h \Psi_R.
\end{equation}
Further, $|\zeta| < |w|$ implies that $|f| \le n$, and similarly
$|\eta| < |w|$ implies that $|h| \le n$.

Suppose $|b_{\eta}| = |b_{\zeta}|$.  Then from (\ref{2nd_star1}) and
(\ref{2nd_star2}) above, $|e| = |g|$.
But $e$ and $g$ are suffixes of $\Psi_R$, so 
we get that $e = g$.  Hence $b_{\zeta} = \Psi_L e = \Psi_L g 
= b_{\eta}$.  Set $b := b_{\zeta} = b_{\eta}$.   
Then from (II) above, as $|b| \le |\Psi_L| + n$, $b$ is a prefix of 
$\Psi_L X_L$.  And from (I) above, $b$ is a suffix of $X_R \Psi_R$.
So $b$ is a non-empty prefix of $w$, and a suffix of $w$.  Hence, as $|b| 
\le \frac{|w|} {2}$, $b$ is a border for $w$.  

So we must have $|b_{\eta}| \neq |b_{\zeta}|$.  Suppose first that 
$|b_{\eta}| < |b_{\zeta}|$.  Now apply Lemma~\ref{loft4} to get that 
there is a positive integer $k$, a non-empty word $p$ and a word $q$ for which 
$\Psi_L = h(pq)^{k}p$ and $\Psi_R = (pq)^{k}pg$.  And finally observe that 
$|pq| + |g| = |f| \le n$. 
If $|b_{\eta}| > |b_{\zeta}|$, the argument is similar, so Claim~\ref{claim2}
is established.

\begin{claim}\label{claim3}
Let $x := pq$ in the statement of Claim~\ref{claim2}.  There is a 
conjugate $c_L$ of $x$ which is a prefix of 
$\Psi_L$, and there is a conjugate $c_R$ of $x$ which is a suffix of 
$\Psi_R$.
\end{claim}

To justify Claim~\ref{claim3}, let $S_L$ be the prefix of length $n$ of
$\Psi_L$.  So there is a word $T_L$ for which we can write
$\Psi_L X_L \theta X_R = S_LT_L$.  (So $w$ is $S_L T_L \Psi_R$.)
Now apply Lemma~\ref{loft3} to $(S_L, T_L, \Psi_R)$, obtaining a word $t_L$, 
with $|t_L| < n$ for which $w_1 := S_L t_L \Psi_R \in L$.  
By supposition, since $N \le |w_1| < |w|$, $w_1$ has a border, say $b_1$.  
Further, if $|b_1| \le n$ then $b_1$ would be a border for $w$.
So we must have $|b_1| >  n$.  And $|b_1| \le \frac{|w_1|} {2}$ implies
$|b_1| \le |\Psi_R|$.  

So $b_1$ is a suffix of $\Psi_R$ of length greater than $n$; hence by
Claim~\ref{claim2} above we can write $b_1 = s_x x^{k_2}pg$ for some integer
$k_2 \ge 0$, with $s_x$ a suffix of $x$.  Write $x = p_xs_x$, and recall that
$p$ is a prefix of $x$.  Then  $|s_x x^{k_2}pg| > n$ and $|x| + |g| \le n$
(from Claim~\ref{claim2}) yields that $s_xp_x$ is a prefix of  
$s_xx^{k_2}pg$, that is, $s_xp_x$ is a prefix of $b_1$.  So set $c_L := 
s_xp_x$.  Since $b_1$ is a prefix of $w_1$,
$c_L$ must be a prefix of $w_1$, and $|c_L| \le n = |S_L|$ gives that
$c_L$ is a prefix of $S_L$, and the first statement of Claim~\ref{claim3}
follows. 

To get the second statement of Claim~\ref{claim3}, similarly 
let $S_R$ be the suffix of length $n$ of $\Psi_R$.
So there is a word $T_R$ for which we can write $X_L \theta X_R \Psi_R =
T_RS_R$.  (So $w$ is $\Psi_L T_R S_R$.)
Now apply Lemma~\ref{loft3} to $(\Psi_L, T_R, S_R)$, obtaining a word $t_R$, with $|t_R| < n$ for
which $w_2 := \Psi_L t_R S_R \in L$.  
By supposition, since $N \le |w_2| < |w|$,  $w_2$ has a border, say $b_2$.  
Further, if $|b_2|
\le n$ then $b_2$ would be a border for $w$.  
So we can assert that $n < |b_2| \le |\Psi_L|$.  

So $b_2$ is a prefix of $\Psi_L$ of length greater than $n$; hence by 
Claim~\ref{claim2} we can write $b_2 = hx^{k_3}\rho_x$ for some integer
$k_3 \ge  0$, with $\rho_x$
a prefix of $x$.  Write $x = \rho_x \sigma_x$.  Then $|hx^{k_3} \rho_x | > n$ 
and $|x| + |h| \le n$ (from Claim~\ref{claim2}) yields that $\sigma_x \rho_x$
is a suffix of $hx^{k_3} \rho_x$, that is, $\sigma_x \rho_x$ is a suffix of
$b_2$.  So  set $c_R := \sigma_x \rho_x$.  Since $b_2$ is a suffix of $w_2$,
$c_R$ must be a suffix of $w_2$, and also $|c_R| \le n = |S_R|$ yields 
that $c_R$ is a suffix of $S_R$, and the second statement of 
Claim~\ref{claim3} follows. 

To complete the proof of Theorem~\ref{loft_thm1}, note that,
since $c_L$ and $c_R$ are both conjugates of $x$,
$c_L$ and $c_R$ are non-empty words which are conjugates.
So there is a non-empty word $\alpha$ and 
a word $\beta$ for which we can write $c_L = \alpha \beta$ and 
$c_R = \beta \alpha$.  Then $\alpha$ is a prefix of 
$\Psi_L$, and $\alpha$ is a suffix of $\Psi_R$, which gives that $\alpha$ 
is a border for $w$, and gives a contradiction.
\end{proof}

\begin{corollary}
The problem of determining if an NFA accepts an unbordered word
is decidable.
\end{corollary}

\begin{proof}
Let $M$ be an NFA with $n$ states.  To determine if $M$ accepts
an unbordered word, it suffices to test whether $M$ accepts
an unbordered word of length at most $6n+1$.
\end{proof}

We do not know if there is a polynomial-time algorithm to
test if an NFA accepts an unbordered word or if the problem is
computationally intractable.

Theorem~\ref{loft_thm1} gives an upper bound of $6n+1$ on the length
of a shortest unbordered word accepted by an $n$-state NFA.  The best
lower bound we are able to come up with is $2n-3$, as illustrated by the
following example: an NFA of $n$ states accepts
$a b^{n-3} a b^*$, and the shortest unbordered word accepted is
$a b^{n-3} a b^{n-2}$, which is of length $2n-3$.  

\begin{theorem}\label{loft_thm2}
Let $M$ be an $n$-state NFA, and let $L$ be the language accepted by $M$.  
Suppose there is an unbordered word in $L$ of length greater than $4n^2 + 6n + 1$.
Then $L$ contains infinitely many unbordered words. 
\end{theorem}
    
\begin{proof}
Suppose $L$ contains only finitely many unbordered words. 
Let $w$ be an unbordered word in $L$ of length greater than $4n^2 + 6n + 1$,
with $|w|$ maximal.  
Write $w$ as $\Psi_L X_L \theta X_R \Psi_R$ for words  $\Psi_L$, $X_L$, 
$\theta$, $\Psi_R$, $X_R$ for which $|X_L| = |X_R| = n$, $|\Psi_L| = 
|\Psi_R| > 2n^2 + 2n$, and $|\theta| \le 1$.    
 We proceed by making the following series of observations.  

\begin{claim}\label{2nd_claim1}
There are words $x_L$, $u_L$, $y_L$ and  $x_R$, $u_R$, $y_R$, 
with $u_L$ and $u_R$ both non-empty, $X_L = x_Lu_Ly_L$, $X_R = x_Ru_Ry_R$, and 
for which:  
\begin{itemize}
\item[(i)] $\zeta := \Psi_L x_Lu_Lu_Ly_L\theta X_R \Psi_R \in L$, and
\item[(ii)] $\eta := \Psi_L X_L \theta x_Ru_Ru_Ry_R \Psi_R \in L$.
\end{itemize}
Further, $|\zeta | > |w|$, and $|\eta| > |w|$.  
\end{claim}

To justify (i), apply Lemma~\ref{loft2} (with $d = 1$) to the 3-tuple $(\Psi_L, X_L, 
\theta X_R \Psi_R)$.  Similarly, to arrive at (ii), apply Lemma~\ref{loft2} again 
(also with $d = 1$) to the 3-tuple $(\Psi_L X_L \theta, X_R, \Psi_R)$.     

\begin{claim}\label{2nd_claim2}
We can write $\Psi_L = h(pq)^{k}p$ and $\Psi_R = (pq)^{k}pg$ 
for $p$ a non-empty word, $g$, $h$ and $q$ words for which $|g| = |h|$, 
$|pq| + |g| \le 2n$, and $k$ an integer $\ge n$.    
Hence $w$ can be written as $h(pq)^{k}p X_L \theta X_R (pq)^{k}pg$. 
\end{claim}

To justify Claim~\ref{2nd_claim2},
first recall that $w = \Psi_L x_Lu_Ly_L \theta x_Ru_Ry_R  
\Psi_R$, and $X_L = x_Lu_Ly_L$, $X_R = x_Ru_Ry_R$.    
From Claim~\ref{2nd_claim1} above and the maximality of $|w|$ 
we get that $\zeta$ and $\eta$ are bordered words, so  
we can assert that there exist non-empty words $b_{\zeta}$ and 
$b_{\eta}$, and words $p_{\zeta}$ and $p_{\eta}$, for which:
\begin{itemize}
\item[(I)]    $\zeta = \Psi_L x_Lu_Lu_Ly_L \theta X_R  
\Psi_R = b_{\zeta} p_{\zeta} b_{\zeta}$, and 
\item[(II)]   $\eta = \Psi_L X_L \theta x_Ru_Ru_Ry_R \Psi_R = 
b_{\eta}p_{\eta} b_{\eta}$.
\end{itemize}

Note that, if  $|b_{\zeta}| \le |\Psi_L|$ then by (I) $b_{\zeta}$ 
would be a border for $w$.  So we must have  $|b_{\zeta}| > |\Psi_L|$.
Similarly, (II) gives that  $|b_{\eta}| > |\Psi_L|$.
These latter facts together with (I) and (II) give that there exists 
non-empty words $e$, $f$, 
$g$, $h$, for which $|e| = |f|$, $|g| = |h|$, and for which
\begin{equation}\label{3rd_star1}
b_{\zeta} = \Psi_L e = f \Psi_R,
\end{equation}
and
\begin{equation}\label{3rd_star2}
b_{\eta} = \Psi_L g = h \Psi_R.
\end{equation}
  
Further, the reader can verify that $|e| \le 2n < |\Psi_R|$, and $|g| \le 2n < |\Psi_R|$. 

Suppose $|b_{\eta}| = |b_{\zeta}|$.  Then from (\ref{3rd_star1}) and (\ref{3rd_star2}) above, 
$|e| = |g|$.  But $e$ and $g$ are suffixes of $\Psi_R$, so 
we get that $e = g$.  Hence $b_{\zeta} = \Psi_L e = \Psi_L g 
= b_{\eta}$.  Set $b := b_{\zeta} = b_{\eta}$.   
Now $|u_Ly_L \theta X_R| > |x_Lu_L|$, so from (I) above, we must have 
$|b| \le |u_Ly_L \theta X_R \Psi_R|$, that is, $b$ is a suffix of $u_Ly_L \theta X_R \Psi_R$. 
Similarly, $|X_L \theta x_Ru_R| > |u_Ry_R|$, so from (II) above we get that
$|b| \le |\Psi_L X_L \theta x_Ru_R|$, that is, $b$ is a prefix of $\Psi_L X_L \theta x_Ru_R$. 
So $b$ is a non-empty prefix of $w$, and a suffix of $w$.
Hence $w$ must be bordered, which is a contradiction. 

So we must have $|b_{\eta}| \neq |b_{\zeta}|$.  First, suppose 
$|b_{\eta}| < |b_{\zeta}|$.
Now apply Lemma~\ref{loft4} to get that 
there is a positive integer $k$, a non-empty word $p$ and a word $q$ for which 
$\Psi_L = h(pq)^{k}p$ and $\Psi_R = (pq)^{k}pg$.  
And finally observe that $|pq| + |g| = |f| \le 2n$, 
and since $|\Psi_L| > 2n^2 + 2n$ and $|pq| \le 2n$, we get that $k \ge n$.
The case $|b_{\eta}| > |b_{\zeta}|$ is symmetric,  so Claim~\ref{2nd_claim2}
is established.     

\begin{claim}\label{2nd_claim3}
Let $x := pq$ in the statement of Claim~\ref{2nd_claim2}.  There is a 
conjugate $c_L$ of $x$ which is a prefix of 
$\Psi_L$, and there is a conjugate $c_R$ of $x$ which is a suffix of 
$\Psi_R$.
\end{claim}

To justify Claim~\ref{2nd_claim3}, recall from Claim~\ref{2nd_claim2}
that $w$ is $\Psi_L X_L \theta X_R x^{k}pg$. 
And since $k \ge n$, we can apply Lemma~\ref{loft2} to the 3-tuple of words
$(\Psi_LX_L \theta X_R, x^k, pg)$, with $d := |x|$, obtaining a 
positive integer $J_1$ for which, for all $z \ge 0$, we have
$\Psi_LX_L \theta X_R x^{k+J_1z}pg \in L$.  
So choose $z_1 := |\Psi_LX_L \theta X_R|$, and define $w_1 :=
\Psi_LX_L \theta X_R x^{k+J_1z_1}pg$.  By supposition $w_1$ is a bordered word, say 
with border $b_1$.  Further, if $|b_1| \le |\Psi_R|$ then $b_1$ would be a border for $w$.  
So we must have $|b_1| > |\Psi_R|$.  And $|b_1| \le \frac{|w_1|} {2}$ implies 
$|b_1| \le |x^{k+J_1z_1}pg|$.  

So $b_1$ is a suffix of $x^{k+J_1z_1}pg$ of length greater than $|\Psi_R| > 2n$, 
hence by Claim~\ref{2nd_claim2} above we can write
$b_1 = s_x x^{k_2}pg$ for some integer $k_2 \ge 0$, 
with $s_x$ a suffix of $x$.  Write $x = p_xs_x$, and recall that $p$ is a 
prefix of $x$.  Then  $|s_x x^{k_2}pg| > 2n$ and $|x| + |g| \le 2n$ (from 
Claim~\ref{2nd_claim2}) yields that $s_xp_x$ is a prefix of  
$s_xx^{k_2}pg$, that is, $s_xp_x$ is a prefix of $b_1$.  So set $c_L := 
s_xp_x$.  Since $b_1$ is a prefix of $w_1$,
$c_L$ must be a prefix of $w_1$, and $|c_L| \le 2n$ gives that
$c_L$ is a prefix of $\Psi_L$, and the first statement of
Claim~\ref{2nd_claim3} follows. 

To justify the second statement of Claim~\ref{2nd_claim3},
we proceed similarly; that is, we recall that
$w$ is $hx^kpX_L \theta X_R \Psi_R$, and 
apply Lemma~\ref{loft2} to the 3-tuple of words
$(h, x^k, pX_L \theta X_R \Psi_R)$, with $d := |x|$, allowing us to assert that there is a
positive integer $J_2$ for which, for all $z \ge 0$, we have
$hx^{k+J_2z}pX_L \theta X_R \Psi_R \in L$.   
So choose $z_2 := |pX_L \theta X_R \Psi_R|$, and define 
$w_2 : = hx^{k+J_2z_2}pX_L \theta X_R \Psi_R $.  By supposition $w_2$ is a bordered word, say 
with border $b_2$.   
Further, if $|b_2| \le |\Psi_L|$ then $b_2$ would be a border for $w$.  So we must have $|b_2| > 
|\Psi_L|$.  And $|b_2| \le \frac{|w_2|} {2}$ implies $|b_2| \le |hx^{k+J_2z_2}p|$.  

So $b_2$ is a prefix of $hx^{k+J_2z_2}p$ of length greater than $|\Psi_L| > 2n$; 
hence by Claim~\ref{2nd_claim2} we can write
$b_2 = hx^{k_3}\rho_x$ for some integer $k_3 \ge 
0$, with $\rho_x$ a prefix of $x$.  Write $x = \rho_x \sigma_x$.  
Then $|hx^{k_3} \rho_x | > 2n$ and $|x| + |h| \le 2n$
(from Claim~\ref{2nd_claim2}) yields that $\sigma_x \rho_x$ is a suffix of
$hx^{k_3} \rho_x$, that is, $\sigma_x \rho_x$ is a suffix of $b_2$.  So 
set $c_R := \sigma_x \rho_x$.  Since $b_2$ is a suffix of $w_2$,
$c_R$ must be a suffix of $w_2$, and also $|c_R| \le 2n$ yields 
that $c_R$ is a suffix of $\Psi_R$, and the second statement of 
Claim~\ref{2nd_claim3} follows. 

To complete the proof of Theorem~\ref{loft_thm2},
note that, since $c_L$ and $c_R$ are 
both conjugates of $x$, $c_L$ and $c_R$ are non-empty 
words which are conjugates.  So there is a non-empty word $\alpha$ and 
a word $\beta$ for which we can write $c_L = \alpha \beta$ and 
$c_R = \beta \alpha$.  Then $\alpha$ is a prefix of 
$\Psi_L$, and $\alpha$ is a suffix of $\Psi_R$, which gives that $\alpha$ 
is a border for $w$, which is a contradiction.  So we're forced to conclude
that $L$ contains infinitely many unbordered words.   
\end{proof}

\begin{theorem}\label{loft_thm3}
Let $M$ be an $n$-state NFA, and let $L$ be the language accepted by $M$.  
Then the following are equivalent:
\begin{enumerate}
\item $L$ contains infinitely many unbordered words. 
\item There is an unbordered word $w$ in $L$, with $4n^2+6n+2 \le |w| \le 8n^2 + 18n + 5$.
\end{enumerate}
\end{theorem}
    
\begin{proof}
(1) $\rightarrow$ (2).  Suppose all words $w \in L$ whose lengths are in 
$[4n^2+6n+2, 8n^2 + 18n + 5]$ are bordered words.
Then by Theorem~\ref{loft_thm1}, (with $N = 4n^2+6n+2$),
we have that any word in $L$ whose length is at least $4n^2+6n+2$ is bordered, i.e., $L$
contains at most finitely many unbordered words.  

(2) $\rightarrow$ (1).  This follows immediately from Theorem~\ref{loft_thm2}.
\end{proof}

\begin{corollary}
The problem of determining if an NFA accepts infinitely many unbordered words
is decidable.
\end{corollary}

\begin{proof}
Let $M$ be an NFA with $n$ states.  To determine if $M$ accepts
infinitely many unbordered words, it suffices to test whether $M$ accepts
an unbordered word $w$, where $4n^2+6n+2 \le |w| \le 8n^2 + 18n + 5$.
\end{proof}

We do not know if there is a polynomial-time algorithm to
test if an NFA accepts infinitely many unbordered words or if the problem is
computationally intractable.

\section{Final remarks}\label{concl}

      In this paper we examined the complexity of checking various
properties of regular languages, such as consisting only of palindromes,
containing at least one palindrome, consisting only of powers, or containing
at least one power.  In each case (except for the unbordered words),
we were able to provide an efficient algorithm or show that the problem
is likely to be hard.  Our results are summarized in the following table.
Here $M$ is an NFA with $n$ states and $t$ transitions.
When $L$ is the language of unbordered words, it is an open problem
to either find polynomial time algorithms to test if 
(a) $L(M) \intersect L = \emptyset$, and (b) $L(M) \intersect L$ is infinite,
or to show the intractability of these problems.

\bigskip
\begin{figure}[H]
\begin{center}
\begin{tabular}{|c|c|c|c|c|}
\hline
     & decide if & decide if & upper bound on & worst-case  \\
$L$  & $L(M) \intersect L = \emptyset$ & $L(M) \intersect L$ & shortest element  & lower bound  \\
     &      & infinite & of $L(M) \intersect L$ & known  \\
\hline
palindromes & $O(n^2+t^2)$ & $O(n^2+t^2)$ & $2n^2-1$ & ${{n^2}\over 2} - 3n+ 5$  \\
\hline
non-palindromes & $O(n^2+tn)$ & $O(n^2+t^2)$ & $3n-1$ & $3n-1$ \\
\hline
$k$-powers       & $O(n^{2k-1} t^k)$ & $O(n^{2k-1} t^k)$ & $kn^k$ &
	$\Omega(n^k)$  \\
($k$ fixed)  & & & &\\
\hline
$k$-powers & PSPACE- & PSPACE- & &  \\
($k$ part of input) & complete & complete & & \\
\hline
non-$k$-powers & $O(n^3 + t n^2)$ & $O(n^3 + t n^2)$ & $3n$ & $(2+{1 \over {2k-2}}) n - O(1)$ \\
\hline
powers & PSPACE- & PSPACE- & $(n+1)n^{n+1}$ & $e^{\Omega(\sqrt{n\log n})}$ \\
& complete & complete & & \\
\hline
non-powers & $O(n^5)$ & $O(n^5)$ & $3n$  & ${5 \over 2} n - 2$\\
\hline
bordered words & $O(n^3 t^2)$ & $O(n^6 t^2)$ & $2n^2 + n- 1$ & $ {{n^2}\over 2} - 6n+ {{43} \over 2}$ \\
\hline
unbordered & decidable & decidable & $6n+1$ & $2n-3$ \\
words & & & & \\
\hline
\end{tabular}
\end{center}
\end{figure}

\section*{Acknowledgments}

The algorithm mentioned in Section~\ref{nn} for testing if an NFA-$\epsilon$
accepts infinitely many words was suggested to us by Timothy Chan.
We would like to thank both him and Jack Zhao for their ideas on this subject.

\bibliography{abbrevs,pal}
\bibliographystyle{new}

\end{document}